%% file: tfhms_global.tex
\RequirePackage{fix-cm}
\RequirePackage{amsmath, bbm}
\documentclass[epj,nopacs]{svjour}
\pdfoutput=1
\smartqed  
\RequirePackage[]{hyperref}
\RequirePackage[numbers,sort&compress]{natbib}

\usepackage{tabularx} 
\usepackage{graphicx}
\usepackage{axodraw2}
\usepackage{amscd}
\usepackage{slashed}
\usepackage{multirow}
\usepackage{float}
\usepackage{amsmath}
\usepackage{bm}
\usepackage{scalerel,amssymb}
\usepackage[caption=false]{subfig}
\RequirePackage[numbers,sort&compress]{natbib}
\def\smelli{{\tt smelli-2.2.0}}

\def\SMsig{{3.4}}
\input{anc/y3inp.tex}
\input{anc/dy3pinp.tex}
\input{anc/dy3inp.tex}
\def\TFHM{{$Y_3$ model}}
\def\DTFHM{{D$Y_3$ model}}
\def\DTFHMp{{D$Y_3^\prime$ model}}
\journalname{Eur. Phys. J. C}

\begin{document}

\title{Global Fits of Third Family Hypercharge Models to Neutral Current
  B-Anomalies and Electroweak Precision
  Observables}

\titlerunning{Global Fits of $Y_3$ Models}        

\author{B.C. Allanach$^1$ 
  \and
  J.\ Eliel Camargo-Molina$^2$ 
  \and
  Joe Davighi$^1$
}



\institute{DAMTP, University of Cambridge, Wilberforce Road, Cambridge, CB3
  0WA, United Kingdom 
\and
  Department of Physics and Astronomy, Uppsala University, Box 516, SE-751 20
  Uppsala, Sweden
}

\date{Received: date / Accepted: date}
\abstract{While it is known that third family hypercharge models can explain the neutral
  current $B-$anomalies, it was hitherto unclear whether the $Z-Z^\prime$
  mixing predicted by such models could simultaneously
  fit electroweak precision observables.
  Here, we perform global fits of several third family hypercharge models to
  a combination of electroweak data 
  and those data pertinent to the neutral current $B-$anomalies.
  While the
  Standard Model is in tension with this combined data set with a $p-$value of
  \SMpval{},
  simple versions of the models (fitting
  two additional parameters each) provide much improved fits.
  The original Third Family Hypercharge Model, for example, 
  has a $p-$value of 
  $\TFHMpval$, with
    $\sqrt{\Delta \chi^2}=6.5\sigma$.}
\maketitle

\section{Introduction \label{sec:intro}}

The neutral current $B-$anomalies (NCBAs) consist of various measurements
in hadronic particle decays which, collectively, are in tension with Standard
Model (SM) predictions. The particular observables displaying such tension often involve 
an effective vertex with an anti-bottom quark, a strange quark, a muon and an
anti-muon, {\em i.e.}\ $(\bar b s)(\mu^+ \mu^-)$, plus the charge conjugated
version.
Observables such  as the ratios of branching ratios
$R_K^{(*)} = BR(B\rightarrow K^{(*)} \mu^+ \mu^-) / BR(B\rightarrow K^{(*)}
e^+ e^-)$ are not displaying the lepton flavour universality (LFU)
property expected of the SM~\cite{Aaij:2017vbb,Aaij:2019wad,Aaij:2021vac}.
Such observables are of particular interest
because much of the theoretical uncertainty in the prediction cancels
in the ratio, leaving the prediction rather precise.
Other NCBA observables display some disparity with SM predictions even when their
larger theoretical uncertainties are taken into account, for example
$BR(B_s \rightarrow \mu^+ \mu^-)$~\cite{Aaboud:2018mst,Chatrchyan:2013bka,CMS:2014xfa,Aaij:2017vad,LHCbtalk}, $BR(B_s \rightarrow \phi \mu^+ \mu^-)$~\cite{Aaij:2015esa,CDF:2012qwd} and
angular distributions of $B\rightarrow K^{(*)} \mu^+ \mu^-$ decays~\cite{Aaij:2013qta,Aaij:2015oid,Aaboud:2018krd,Sirunyan:2017dhj,Khachatryan:2015isa,Bobeth:2017vxj}.
Global fits find that new physics contributions to the
$(\bar b s)(\mu^+ \mu^-)$
effective vertex can fit the
NCBAs much better than the SM can~\cite{Alguero:2019ptt,Alok:2019ufo,Ciuchini:2019usw,Aebischer:2019mlg,Datta:2019zca,Kowalska:2019ley,Arbey:2019duh}.

A popular option for beyond the SM (BSM) explanations of the NCBAs
is that of a $Z^\prime$
vector boson with family dependent interactions~\cite{Gauld:2013qba,Buras:2013dea,Buras:2013qja,Buras:2014yna,Allanach:2015gkd}. Such a particle is predicted
by models with a BSM spontaneously broken $U(1)$ gauged flavour
symmetry.
The additional quantum numbers of the SM fermions are constrained by the need
to cancel local
anomalies~\cite{Ellis:2017nrp,Allanach:2018vjg,Allanach:2020zna}, 
for example muon minus tau lepton
number~\cite{Altmannshofer:2014cfa,Crivellin:2015mga,Crivellin:2015lwa,Crivellin:2015era,Altmannshofer:2015mqa,Davighi:2020qqa},
third family baryon number minus second family lepton
number~\cite{Alonso:2017uky,Bonilla:2017lsq,Allanach:2020kss}, third family
hypercharge~\cite{Allanach:2018lvl,Davighi:2019jwf,Allanach:2019iiy} or other
assignments~\cite{Sierra:2015fma,Celis:2015ara,Greljo:2015mma,Falkowski:2015zwa,Chiang:2016qov,Boucenna:2016wpr,Boucenna:2016qad,Ko:2017lzd,Alonso:2017bff,Tang:2017gkz,Bhatia:2017tgo,Fuyuto:2017sys,Bian:2017xzg,King:2018fcg,Duan:2018akc,Kang:2019vng,Calibbi:2019lvs,Altmannshofer:2019xda,Capdevila:2020rrl}. 

The current paper is about the third family hypercharge option.
The Third Family Hypercharge Model~\cite{Allanach:2018lvl}
({henceforth abbreviated as the `\TFHM{}'}) explains the hierarchical heaviness
of the third family and the smallness of quark mixing.
It was shown to successfully fit NCBAs,
along with constraints from $B_s-{\bar B}_s$
mixing and LFU constraints on $Z^0$ boson interactions.
The ATLAS experiment at the LHC has searched\footnote{During the final stages
  of preparation of this manuscript, the CMS experiment released a di-muon resonance search~\cite{Sirunyan:2021khd}
  with similar exclusion regions.} for the production of BSM resonances that yield a
peak in the di-muon invariant mass ($m_{\mu \mu}$) spectrum, but have yet to find a
significant one~\cite{Aad:2019fac}. This implies a lower bound upon the mass of the
$Z^\prime$ in the \TFHM{}, $M_X>1.2$ TeV~\cite{Allanach:2019mfl},
 but
plenty of viable parameter space remains which successfully explains the
NCBAs.
A variant, the Deformed Third Family Hypercharge Model (\DTFHM{}),  was subsequently
introduced~\cite{Allanach:2019iiy} {in order to remedy a somewhat} ugly
feature (ugly from a naturalness point of view) in the
construction of the {original \TFHM{}: namely, that} a Yukawa coupling allowed at the renormalisable
level was assumed to be tiny in order to agree with  strict lepton flavour
violation constraints. The \DTFHM{} can simultaneously fit the NCBAs and be consistent with
the ATLAS di-muon direct search constraint for 1.2 TeV$< M_X<$12
TeV. We will also present results for a third variant, the \DTFHMp{}, which is
identical to the \DTFHM{} but with charges for second and third family leptons
interchanged. As we show later on, this results in a better fit to data due to
the different helicity structure of the couplings of the $Z^\prime$ boson to muons (see section~\ref{sec:DTFHMresults} for details).

In either of these third family hypercharge models, the local gauge symmetry of the SM
is\footnote{{Possible} quotients {of the gauge group} by discrete subgroups play no role in our argument and
  so we shall {henceforward} ignore them.}
extended to $SU(3)\times SU(2) \times U(1)_Y \times U(1)_X$. {This is}
spontaneously broken to the SM gauge group by {the non-zero vacuum expectation value (VEV) of} a SM-singlet `flavon' field $\theta$ that has
a non-zero $U(1)_X$ charge. In each model, the third family quarks' $U(1)_X$ charges
are equal to their hypercharges whereas the first two family quarks are
chargeless under $U(1)_X$. We must (since it is {experimentally determined to
  be} ${\mathcal O}(1)$ and is therefore inconsistent with a suppressed,
non-renormalisable coupling)
ensure that a renormalisable top Yukawa coupling is allowed by
$U(1)_X$; this implies that the SM Higgs doublet field should have {$U(1)_X$} charge equal to
its hypercharge. 
{Consequently,} when the Higgs doublet acquires a VEV to break the
electroweak symmetry, {this gives rise to $Z^0-Z^\prime$ mixing~\cite{Allanach:2018lvl}.} Such mixing is subject to stringent
constraints from electroweak precision observables (EWPOs), in particular {from} the
$\rho$-parameter, which encodes the ratio of the masses of the $Z^0$ boson and
the $W$ boson~\cite{Davighi:2020nhv}. 

Third family hypercharge models can fit the NCBAs for a range of the ratio of the $Z^\prime$ gauge
coupling to its mass $g_{X}/M_{X}$ which does {\em not}\/ contain zero. This means that it is not possible to `tune the $Z-Z^\prime$
mixing away' if one wishes
the model to fit the NCBAs. As a consequence, 
it is not clear whether the EWPOs will strongly preclude the {(D)\TFHM{}s} from
explaining the NCBAs or not.

The purpose of this paper is to perform a global fit to a combined set of
electroweak and NCBA-type data, along with other relevant constraints on
flavour changing neutral currents (FCNCs). It is clear that the SM provides a
poor fit to this combined set, as Table~\ref{tab:sm_fit} shows. A
$p-$value\footnote{All $\chi^2$ and $p-$values which we present here are
  calculated in \smelli{}. 
  We estimate that the numerical uncertainty in the \smelli{} calculation
  of a global
  $\chi^2$ value is $\pm 1$ and the resulting   uncertainty in the second
  significant figure of any global $p-$value quoted is 
  $\pm   3$.} of 
\SMpval{} 
corresponds to `tension at the $\SMsig{}\sigma$ level'.
\begin{table}
  \begin{center}
    \begin{tabular}{|c|ccc|} \hline
      data set & $\chi^2$ & $n$ & $p-$value \\ \hline
      \input{anc/sm_fit.tex} \hline
    \end{tabular}
    \caption{\label{tab:sm_fit} SM goodness of fit for the different data sets
      we consider, as calculated by \smelli{}. We display the total Pearson's chi-squared
      $\chi^2$ for each data set along with the number of observables $n$ and
      the data set's $p-$value. The set named `quarks' contains $BR(B_s
      \rightarrow \phi \mu^+ \mu^-)$, 
      $BR(B_s\rightarrow \mu^+\mu^-)$,  $\Delta m_s$ and various differential
      distributions in $B\rightarrow K^{(\ast)} \mu^+\mu^-$ decays among others, whereas `LFU
      FCNCs' contains $R_{K^{(\ast})}$ and some $B$ meson decay branching ratios
      into di-taus. Our sets are identical to those defined by \smelli{}
      and we refer the curious reader to its manual~\cite{Aebischer:2018iyb},
      where the observables are enumerated.     We have updated $R_K$ and $BR(B_{s,d}\rightarrow \mu^+ \mu^-)$ with the
    latest LHCb measurements as detailed in the text.
}
  \end{center}
\end{table}
The (D)\TFHM{}s are of particular interest as plausible models of new physics
if they fit the data significantly better than the SM\@,
a question which can best be settled by performing appropriate global fits.

Our paper proceeds as follows:
we introduce the models and define their parameter spaces in
\S\ref{sec:models}.
At renormalisation scales at or below $M_{X}$ but above $M_W$,
we encode the new physics effects in each model via 
the Standard Model Effective Theory (SMEFT). We calculate the leading (dimension-6) SMEFT operators
predicted by our models at the scale $M_{X}$ in \S\ref{sec:wcs}. These
provide the input to the 
calculation of observables by \smelli{}~\cite{Aebischer:2018iyb}\footnote{We
  use the development version of \smelli{} that was released on {\tt github}
  on 8$^{th}$ March 2021 {which we have updated to take into account 2021 LHCb
    measurements of $R_K$, $BR(B_s \rightarrow \mu^+ \mu^-)$ and
    $BR(B\rightarrow \mu^+ \mu^-)$.}}, which we
describe {at the beginning of \S\ref{sec:fits}}. The results of the fits are presented in
{the remainder of \S\ref{sec:fits},} before a discussion in \S\ref{sec:disc}.

\section{Models \label{sec:models}}
In this section we review
the models of interest to this study, in sufficient detail
so as to proceed with the
calculation of the SMEFT Wilson coefficients (WCs) in the following section.
Under $SU(3)\times SU(2) \times U(1)_Y$, we define the fermionic
fields such
that they transform in the
following representations:
${Q_L}_i:=({u_L}_i\ {d_L}_i)^T \sim (\bm{3},\ \bm{2},\ +1/6)$,
${L_L}_i:=({\nu_L}_i\ {e_L}_i)^T \sim (\bm{1},\ \bm{2},\ -1/2)$,
${e_R}_i \sim (\bm{1},\ \bm{1},\ -1)$,
${d_R}_i \sim (\bm{3},\ \bm{1},\ -1/3)$,
${u_R}_i \sim (\bm{3},\ \bm{1},\ +2/3)$, where $i\in \{1,2,3\}$ is a family index ordered by
increasing mass. Implicit in the definition of these fields is that we have
performed a flavour rotation so that ${d_L}_i, {e_L}_i, {e_R}_i,
{d_R}_i, {u_R}_i$ are mass basis fields.
In what follows,  we denote 3-component
column vectors in family space with bold font, for example
${\bf u_L}:=(u_L,\ c_L,\ t_L)^T$.
The Higgs doublet is a complex scalar $\phi \sim (\bm{1},\ \bm{2},\ +1/2)$,
and all three models which we consider ({the \TFHM{}, the \DTFHM{} and the \DTFHMp{}})
incorporate a complex scalar flavon with SM quantum numbers $\theta \sim (\bm{1}, \bm{1}, 0)$, which has a
$U(1)_X$ charge $X_\theta \neq 0$ and is used to Higgs the {$U(1)_X$} symmetry, such that its gauge
boson acquires a mass at the TeV scale or higher.

In the following we will present our results for three variants of third family hypercharge models, which differ in the charge assignment for the SM fields:
\begin{itemize}
\item The \TFHM{}, introduced in \cite{Allanach:2018lvl}. Only third
  generation fermions have non-zero $U(1)_X$ charges. The charge assignments can be read in Table~\ref{tab:chargesTFHM}.
\item The \DTFHM{} as introduced in \cite{Allanach:2019iiy}. It differs from the \TFHM{} in that charges have been assigned to the second generation leptons as well, while still being anomaly free. 
\item The \DTFHMp{}, which differs from the \DTFHM{} in that the charges for
  third and second generation left-handed leptons are interchanged. The charge
  assignments can be read in Table~\ref{tab:chargesDTFHMp}.
\end{itemize}
All three of these gauge symmetries have identical couplings to quarks,
  coupling only to the third family via hypercharge quantum numbers. This
  choice means that, of the quark Yukawa couplings, only the top and
  bottom Yukawa couplings are present at the renormalisable
  level. Of course, the light quarks are not massless in reality; their
  masses, as well as the small quark mixing angles, must be encoded in
  higher-dimensional operators that come from a further layer of heavy
  physics, such as a suite of heavy vector-like fermions at a
  mass scale $\Lambda > M_X/g_X$, where $g_X$ is the $U(1)_X$ gauge coupling.

Whatever this heavy physics might be, the structure of the
  light quark Yukawa couplings will be governed by the size of   parameters that break the $U(2)^3_\mathrm{global}:=U(2)_q \times U(2)_u
  \times U(2)_d$ accidental global
  symmetry~\cite{Pomarol:1995xc,Barbieri:1995uv,Barbieri:2011ci,Blankenburg:2012nx,Barbieri:2012uh}
  of the renormalisable third family hypercharge lagrangians. For example, a
  minimal set of spurions charged under both $U(2)^3_\mathrm{global}$ and
  $U(1)_X$ was considered\footnote{In the present paper we take a more
      phenomenological approach in specifying the fermion mixing matrices, but
      nonetheless, the quark mixing matrices that we use are qualitatively similar
      to those expected from the $U(2)^3_\mathrm{global}\times U(1)_X$
      breaking spurion analysis. For example, there is no mixing in the
      right-handed fields.} in Ref.~\cite{Davighi:2021oel}, which reproduces
  the observed hierarchies in quark masses and mixing angles when the scale
  $\Lambda$ of new physics is a factor of $15$ or so larger than
  $M_X/g_X$. Taking this hierarchy of scales as a general guide, and observing
  that the global fits to electroweak and flavour data that we perform in this
  paper prefer $M_X/g_X\approx 10$ TeV, we expect the new physics
  scale to
   be around $\Lambda
  \approx 150$ TeV. This scale is high enough to suppress most contributions
  of the heavy physics, about which we remain agnostic, to low energy
  phenomenology including precise flavour bounds\footnote{Kaon mixing is one possible exception
  where such heavy physics might play a non-trivial role, however.}. For this reason, we feel safe in neglecting the
  contributions of the $\Lambda$ scale physics to the SMEFT coefficients that
  we calculate in \S~\ref{sec:wcs}, and shall not consider it in any further detail. 

Continuing, we will first detail the scalar sector, which is common to (and identical in) all of the
third family hypercharge models, before going on to discuss aspects of each model that are different (most importantly, the couplings to leptons).

\subsection{The scalar sector}

The coupling of the flavon to the $U(1)_X$ gauge field is encoded in the covariant derivative
\begin{equation}
  D_\mu \theta = (\partial_\mu + i X_\theta g_{X} X_\mu) \theta,
\end{equation}
where $X_\mu$ is the
$U(1)_X$ gauge boson in the unbroken phase and $g_{X}$ is its gauge coupling.
The flavon $\theta$ is assumed to acquire a VEV $\langle \theta \rangle$ at (or above) the TeV scale, which
spontaneously breaks $U(1)_X$. Expanding
$\theta = (\langle \theta \rangle + \vartheta)/\sqrt{2}$, its kinetic terms $(D_\mu \theta)^\dag D^\mu \theta$ in the Lagrangian density
give the gauge boson a mass $M_{X}=X_\theta g_{X} \langle \theta
\rangle$ through the Higgs mechanism. After electroweak symmetry breaking, 
the electrically-neutral gauge bosons $X$, $W^3$ and $B$ mix,  
giving rise to $\gamma$, $Z^0$ and $Z^\prime$ as the physical mass eigenstates~\cite{Allanach:2018lvl}. To terms of order $(M_Z^2 /
  M_{X}^2)$, the mass and the couplings of the $X$ boson are equivalent to
  those of the $Z^\prime$ boson. Because we take $M_X \gg M_Z$, the matching to the SMEFT (\S\ref{sec:wcs}) 
should be done in the unbroken electroweak phase, where it is the $X$ boson that is properly integrated out. In the rest of this section we therefore 
specify the $U(1)_X$ sector via the $X$ boson and its couplings. Throughout this paper, we entreat the reader
to bear in mind that in terms of searches and several other aspects of their phenomenology,
to a decent approximation the $X$ boson and the
    $Z^\prime$ boson are synonymous. 

\sloppy The covariant derivative of the Higgs doublet is 
\begin{equation}
  D_\mu \phi = \left(\partial_\mu + i \frac{g}{2} \sigma^a W_\mu^a + i
  \frac{g^\prime}{2} B_\mu + i \frac{g_X}{2} X_\mu\right) \phi,
\end{equation}
 where $W_\mu^a$ ($a=1,2,3$) are unbroken $SU(2)$ gauge bosons, $\sigma^a$ are the Pauli
matrices, $g$ is the $SU(2)$ gauge coupling, $B_\mu$ is the hypercharge gauge
boson and $g^\prime$ is the hypercharge gauge coupling. 
The kinetic term for the Higgs field, $(D_\mu \phi)^\dagger (D^\mu \phi)$, contains terms both linear and 
quadratic in $X_\mu$. It is the linear terms
\begin{equation} \label{eq:Higgs_UV}
\mathcal{L} \supset -i \frac{g_X}{2} X_\mu \phi^\dagger  \left(\partial_\mu + i \frac{g}{2} \sigma^a W_\mu^a + i
  \frac{g^\prime}{2} B_\mu \right) \phi + \text{h.c.}
\end{equation}
that, upon integrating out the $X_\mu$ boson, will give the leading contribution to 
the SMEFT in the form of dimension-6 operators involving the Higgs, as we describe in \S\ref{sec:wcs}.

The charges of the fermion fields differ between the \TFHM{} and the \DTFHMp{}, as follows.

\subsection{Fermion couplings: the \label{sec:TFHMdef} \TFHM{}}

\begin{table}
\begin{center}
\begin{tabular}{|ccc|}\hline
$X_{Q_i}=0$ &  $X_{{u_R}_i}=0$ & $X_{{d_R}_i}=0$ \\ 
$X_{Q_3}=1/6$ &  $X_{{u_R}_3}=2/3$ & $X_{{d_R}_3}=-1/3$ \\
$X_{L_i}=0$ & $X_{{e_R}_i}=0$ & $X_{{\phi}}=1/2$ \\
$X_{L_3}=-1/2$ & $X_{{e_R}_3}=-1$ & $X_\theta$ \\ 
\hline
\end{tabular}
\caption{\label{tab:chargesTFHM} $U(1)_X$ charges of the gauge eigenbasis fields in the \TFHM{}, where $i \in \{1, 2\}$. The flavon charge $X_\theta$ is left undetermined.}
\end{center}
\end{table}

\sloppy The \TFHM{} has fermion charges as listed in Table~\ref{tab:chargesTFHM} (in the gauge eigenbasis), leading to the following Lagrangian density describing the $X$ boson-SM fermion couplings~\cite{Allanach:2018lvl}: 
\begin{eqnarray}
  {\mathcal L}_{{Y_3}}^\psi
  =-&g_{X}&\left( 
  \frac{1}{6}\overline{\bf Q_L} \Lambda^{(d_L)}_\xi \slashed{X} {\bf Q_L}-
  \frac{1}{2}\overline{\bf L_L} \Lambda^{(e_L)}_\xi \slashed{X} {\bf L_L}
  \right. \nonumber \\ && 
  + \frac{2}{3}
  \overline{\bf u_R} \Lambda^{(u_R)}_\xi \slashed{X} {\bf u_R}-
  \frac{1}{3}
  \overline{\bf d_R} \Lambda^{(d_R)}_\xi \slashed{X} {\bf d_R}\nonumber \\ && \left.
  -\overline{\bf e_R} \Lambda^{(e_R)}_\xi \slashed{X} {\bf e_R}
  \right),  
  \label{ZpcoupTFHM}  
\end{eqnarray}
where
\begin{equation}
  \Lambda^{(I)}_P := V_{I}^\dagger P V_{I} 
  \label{lambdas}
\end{equation}
are Hermitian 3-by-3 matrices. {The index}
$I \in \{u_L, d_L, e_L, \nu_L, u_R, d_R, e_R \}$ and {the matrix}  $P \in \{\xi,
\Omega, \Psi\}$, where 
\begin{equation}
  \xi = \left(\begin{array}{ccc}
    0 & 0 & 0 \\ 0 & 0 & 0 \\ 0 & 0 & 1 \\
  \end{array}\right),
\end{equation}
and $\Omega$ and $\Psi$ are described in \S\ref{sec:dtfhm}.
The $V_I$ are 3-by-3 unitary matrices describing the mixing between 
fermionic gauge eigenstates and their mass eigenstates.
Note that the quark doublets have been family rotated so that
  the $d_{L_i}$ (but not the
  $u_{L_i}$ fields) correspond to their mass eigenstates.
  Similarly, we have rotated $L_i$ such that $e_{L_i}$ align with the charged
  lepton   mass eigenstates, but $\nu_{L_i}$ are not.
  This will simplify the matching to the SMEFT operators that we perform
  in \S\ref{sec:wcs}. 
We now go on to cover the $X$ boson couplings in the \DTFHMp{} before detailing
the fermion mixing ansatz (which is common to all three models).

\subsection
    {Fermion couplings: the \DTFHMp{}\label{sec:dtfhm}}

 \begin{table}
\begin{center}
\begin{tabular}{|ccc|}\hline
$X_{Q_1}=0$ &  $X_{{u_R}_1}=0$ & $X_{{d_R}_1}=0$ \\
$X_{Q_2}=0$ &  $X_{{u_R}_2}=0$ & $X_{{d_R}_2}=0$ \\
$X_{Q_3}=1/6$ &  $X_{{u_R}_3}=2/3$ & $X_{{d_R}_3}=-1/3$ \\
$X_{L_1}=0$ & $X_{{e_R}_1}=0$ & $X_{{\phi}}=1/2$  \\
$X_{L_2}=-4/3$ & $X_{{e_R}_2}=2/3$ & $X_\theta$\\
$X_{L_3}=5/6$ & $X_{{e_R}_3}=-5/3$ & 
  \\ \hline
\end{tabular}
\caption{\label{tab:chargesDTFHMp} $U(1)_X$ charges of the gauge eigenbasis fields in the \DTFHMp{}. The original \DTFHM{} charges (as introduced in Ref.~\cite{Allanach:2019iiy}) can be obtained by interchanging $X_{L_3}$ and $X_{L_2}$. The flavon charge $X_\theta$ is left undetermined.}
\end{center}
\end{table}

    For the \DTFHMp{} with the charge assignments listed in
    Table~\ref{tab:chargesDTFHMp}, the Lagrangian contains the following $X$ boson-SM fermion couplings~\cite{Allanach:2019iiy}:
    \begin{eqnarray}
      {\mathcal L}_{{\text{D}Y_3^\prime}}^\psi =
      &-g_{X} &\left( 
      \frac{1}{6}{\overline{{\bf Q_L}}} \Lambda^{(d_L)}_\xi \slashed{X} {\bf
        Q_L} -
      \frac{4}{3} {\overline{{\bf L_L}}} \Lambda^{(e_L)}_\Omega \slashed{X}
           {\bf L_L}
           \right. \nonumber \\ && \left.
           +\frac{2}{3}{\overline{{\bf u_R}}} \Lambda^{(u_R)}_\xi \slashed{X} {\bf u_R}
           -\frac{1}{3} {\overline{{\bf d_R}}} \Lambda^{(d_R)}_\xi \slashed{X} {\bf d_R}
           \right. \nonumber \\ && \left.
           +\frac{2}{3} {\overline{{\bf e_R}}} \Lambda^{(e_R)}_\Psi \slashed{X} {\bf e_R}
           \right). \label{ZpcoupDTFHM}
    \end{eqnarray}
The matrices $\Lambda^{(I)}_{\Omega}$ and $\Lambda^{(I)}_{\Psi}$ are defined in
    (\ref{lambdas}), where
    \begin{equation}
      \Omega= \left( \begin{array}{ccc}
        0 & 0 & 0 \\
        0 & 1 & 0 \\
        0 & 0 & -\frac{5}{8} \\
      \end{array}\right), \qquad    
      \Psi= \left( \begin{array}{ccc}
        0 & 0 & 0 \\
        0 & 1 & 0 \\
        0 & 0 & -\frac{5}{2} \\
      \end{array}\right). 
    \end{equation}
    
    \subsection{Fermion mixing ansatz}

        The CKM matrix and the PMNS matrix are predicted to be
        \begin{equation}
          V=V_{u_L}^\dag V_{d_L}, \qquad U = U_{\nu_L}^\dag V_{e_L}, \label{mixing}
        \end{equation}
        respectively. {For {all of the third family hypercharge models that
          we address here}, the matrix element
        $(V_{(d_L)})_{23}$ must be non-zero} in order
        to obtain new physics contributions of the
        sort required to explain the NCBAs. {Moreover,} in the \TFHM{} we need $(V_{e_L})_{23} \neq 0$ in order to generate a coupling (here left-handed) to muons\footnote{The \DTFHM{} (\DTFHMp{}) does not require
        $(V_{e_L})_{23} \neq 0$ to fit the NCBAs, since it already possesses a
        coupling to $\mu_L$.}. These will lead to a BSM contribution to a
        Lagrangian density term in the weak effective theory proportional to
        $(\bar b \gamma^\mu P_L s)(\bar \mu \gamma_\mu P_L \mu)$, where $P_L$ is the
        left-handed projection operator, which previous fits to the weak effective
        theory indicate is essential in order to fit the NCBAs~\cite{Alguero:2019ptt,Alok:2019ufo,Ciuchini:2019usw,Aebischer:2019mlg,Datta:2019zca,Kowalska:2019ley,Arbey:2019duh}.  
        
\sloppy In order to investigate the model further phenomenologically, we must
assume a particular ansatz for the unitary fermion mixing matrices $V_I$.
Here, for $V_{d_L}$, we choose the `standard parameterisation' often used for the CKM
matrix~\cite{Zyla:2020zbs}. This is a parameterisation of a family of
unitary 3 by 3 matrices that depends only upon one complex phase and three
mixing angles (a more general parameterisation 
would also depend upon five additional complex phases):
        \begin{equation}
          \left(\begin{array}{ccc}
            c_{12}c_{13} & s_{12}c_{13} & s_{13} e^{-i\delta} \\
            -s_{12}c_{23}-c_{12}s_{23}s_{13}e^{i \delta} & c_{12}c_{23} -
            s_{12}s_{23}s_{13}e^{i\delta} & s_{23}c_{13} \\
            s_{12}s_{23}-c_{12}c_{23}s_{13}e^{i \delta} & -c_{12}s_{23} -
            s_{12}c_{23}s_{13}e^{i\delta} & c_{23}c_{13} \\    
          \end{array} \right),
        \end{equation}
        where $s_{ij}:=\sin \theta_{ij}$ {and} $c_{ij}:=\cos \theta_{ij}$, {for angles}
        $\theta_{ij},\ \delta \in
        \mathbb{R}/(2 \pi \mathbb{Z})$.
        To define our {particular} ansatz, we choose angles such that $\left(V_{d_L}\right)_{ij} = V_{ij}$ for all $ij\neq 23$, {\em i.e.}\  we insert the current world-averaged
        measured central values of 
        $\theta_{ij}$ and $\delta$~\cite{Zyla:2020zbs}, {\em except}\/ for the crucial
        mixing angle
        $\theta_{23}$, upon which the NCBAs sensitively
        depend. Thus, we fix the angles and phase such that $s_{12}=0.22650$, $s_{13}=0.00361$ and
        $\delta=1.196$ but allow $\theta_{23}$ to float {as a free parameter
          in our global fits.}  
        Following Refs.~\cite{Allanach:2018lvl,Allanach:2019iiy},
        we choose simple forms for the other mixing matrices
        which are likely to evade strict FCNC bounds. {Specifically, we choose}
        $V_{d_R}=1$, $V_{u_R}=1$ and 
        \begin{equation}
          V_{e_L} = \left(\begin{array}{ccc}
            1 & 0 & 0 \\ 0 & 0 & 1 \\ 0 & 1 & 0 \\
          \end{array}\right) \label{VeR}
        \end{equation} 
		in the \TFHM{}\footnote{Note that in the \TFHM{}, this is
                    equivalent to instead setting $V_{e_L}=1$ and switching the $U(1)_X$
                charges of $L_2$ and $L_3$ in Table~\ref{tab:chargesTFHM}.},
                and $V_{e_L}=1$ in the \DTFHMp{}. Finally, 
        $V_{u_L}$ and $V_{\nu_L}$ are then fixed
		by (\ref{mixing}) and the measured CKM/PMNS matrix entries. For the remainder of this paper,
        when referring to the \TFHM{} or the \DTFHMp{},
        we shall implicitly refer to the versions given by
        this mixing ansatz (which, {we emphasise,} is taken to be the same
        for all third family hypercharge models aside
        from the assignment of $V_{e_L}$).

Next, we turn to calculating the complete set of dimension-6 WCs in the SMEFT that result from integrating the
$X$ boson out of the theory.

\section{SMEFT Coefficients \label{sec:wcs}}
           
So far, particle physicists have found scant direct evidence of new physics below the TeV scale. This motivates the study of BSM models whose
  new degrees of freedom reside at the TeV 
  scale or higher.
In such scenarios, it makes sense to consider the Standard Model as an
effective field theory realisation of the underlying high energy model. If
  one wishes to remain agnostic about further details of the high energy
  theory, this amounts to including all possible operators consistent with
the SM gauge symmetries and performing an expansion in powers of the ratio
of the electroweak and new physics scales. 

\sloppy The Standard Model Effective Field Theory
\cite{Buchmuller:1985jz,Brivio:2017vri,Dedes:2017zog} is such a
parameterisation of the effects of heavy fields beyond the SM (such as a heavy
$X$ boson field of interest to us here) through $d>4$ operators built out of
the SM fields. In this paper we will work with operators up to dimension 6
({\em i.e.}\/ we will go to second order in the power expansion). Expanding the SMEFT to this order gives us a very good approximation to all of
the observables we consider; the relevant expansion parameter for the EWPOs is $(M_Z / M_{X})^2 \ll 1$, and in the case of observables derived from the decay of a meson of mass $m$, the relevant expansion parameter is $(m /
M_{X})^2 \ll 1$.
By restricting to the $M_{X}>1$ TeV region, 
we ensure that both of these mass ratios are small enough to yield a good approximation. 

The SMEFT Lagrangian can be expanded  as
\begin{equation}
\mathcal{L}_{\text{SMEFT}} = \mathcal{L}_{SM} + C_{\text{5}} O_{\text{5}} +
\sum_{\text{dim 6}} C_i O_i + \ldots,
\end{equation}
where $O_{\text{5}}$ schematically indicates Weinberg operators with various flavour indices, which result in neutrino masses and may be obtained by adding heavy gauge singlet chiral fermions to play the role of right-handed neutrinos.
The sum that is explicitly notated then runs over all
mass dimension-6 SMEFT operators, and the ellipsis represents terms  which are of
mass dimension (in the fields) greater than 6. The {WCs} $C_i$ have units of
$[\mbox{mass}]^{-2}$. In the following we shall work in the Warsaw basis,
which defines a basis in terms of a set of independent baryon-number-conserving
operators~\cite{Grzadkowski:2010es}.  By performing
the matching between our models and the SMEFT, we shall obtain the set of WCs $C_i$ at the scale $M_{X}$, which can then be used to calculate predictions for observables.

            \begin{figure}
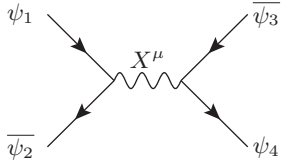

      \begin{center}
        \begin{axopicture}(50,50)(-40,0)
          \Line[arrow](-40,50)(-15,25)
          \Line[arrow](-15,25)(-40,0)
          \Line[arrow](35,50)(10,25)
          \Line[arrow](10,25)(35,0)
          \Photon(-15,25)(10,25){3}{3}
          \Text(-2.5,33)[c]{$X^\mu$}
          \Text(-50,50)[c]{$\psi_1$}
          \Text(-50,0)[c]{$\overline{\psi_2}$}
          \Text(42,50)[c]{$\overline{\psi_3}$}
          \Text(42,0)[c]{$\psi_4$}
        \end{axopicture}
      \end{center}
      \caption{\label{fig:fermions} $X$-boson mediated process responsible
        for the effective vertex between 4 fermionic fields {$\{\psi_i\}$}. }
    \end{figure}

To see where these dimension-6 operators come from, let us first consider the origin of four-fermion operators. We may write the fermionic couplings of the underlying theory {of the $X$ boson}, given in (\ref{ZpcoupTFHM}) and
    (\ref{ZpcoupDTFHM}) for the \TFHM{} and \DTFHMp{} respectively, as 
    \begin{equation}
      {\mathcal L}^\psi = - J^\mu_\psi X_\mu,
    \end{equation}
    where 
    \begin{equation}
      J^\mu_\psi = \sum_{\psi_{{i,j}}} {\kappa_{ij}} \overline{\psi_i} \gamma^\mu \psi_{{j}}
      \end{equation}
    is the fermionic current 
    that the $X$ boson couples to, where the sum runs over all pairs of SM Weyl
    fermions $\psi_i$.  The couplings $\kappa_{ij}$ are identified from (\ref{ZpcoupTFHM}) or
    (\ref{ZpcoupDTFHM}),
    depending upon the model.
    After integrating out
    the $X$ boson in processes such as the one in Fig.~\ref{fig:fermions}, one obtains the following terms in the effective Lagrangian:
    \begin{equation}
	{\mathcal L}_\text{SMEFT} \supset 
	- \frac{{J_\psi}_\mu {J_\psi}^\mu}{2 M_{X}^2}.
    \end{equation}
    We match the terms thus obtained with the four-fermion
    operators in the Warsaw basis~\cite{Grzadkowski:2010es} in order to
    identify the 4-fermion SMEFT WCs in that basis.

\begin{figure}[h]
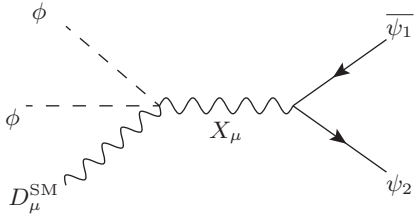

\begin{center}
\begin{axopicture}(200,100)(0,0)
\DashLine(40,80)(75,50){5}
\DashLine(25,50)(75,50){5}
\Photon(40,20)(75,50){3}{5}
\Text(30,85)[c]{$\phi$}
\Text(20,45)[c]{$\phi$}
\Text(28,15)[c]{$D_\mu^{\text{SM}}$}
\Photon(75,50)(125,50){3}{5}
\Text(100,40)[c]{$X_\mu$}
\Line[arrow](160,75)(125,50)
\Line[arrow](125,50)(160,25)
\Text(165,80)[c]{$\overline{\psi_1}$}
\Text(165,20)[c]{$\psi_2$}
\end{axopicture}
\end{center}
\caption{$X$-boson mediated process responsible for the effective vertex
  between two Higgs fields, one SM gauge boson, and a fermion bi-linear operator.
\label{fig:2Higgs_fermions}}    
\end{figure}

These 4-fermion operators are not the only SMEFT operators that are produced at dimension-6 by integrating out the $X$ bosons of our models.
Due to the tree-level $U(1)_X$ charge of the SM Higgs, there are also various operators in the Higgs sector of the SMEFT, as follows. The (linear) couplings of the $X$ boson
to the Higgs, as recorded in (\ref{eq:Higgs_UV}), can again be written as the coupling of $X_\mu$ to a current, {\em viz.}
    \begin{equation}
      {\mathcal L}^\phi = - {J_\phi}_\mu X^\mu,
    \end{equation}
    where this time
    \begin{equation}
      {J_\phi}_\mu =  i \frac{g_X}{2} \phi^\dagger D_\mu^\text{SM} \phi + \text{h.c.}
      \end{equation}
is the bosonic current to which the $X$ boson couples, where $D_\mu^\text{SM}=\partial_\mu + i \frac{g}{2} \sigma^a W_\mu^a + i
  \frac{g^\prime}{2} B_\mu$.
Due to the presence of $X$ boson couplings to operators which are bi-linear in both the fermion
fields ($J_\psi^\mu$) and the Higgs field ($J_\phi^\mu$), integrating out the $X$ bosons gives rise to cross-terms 
    \begin{equation}
      {\mathcal L}_\text{SMEFT} \supset 
	- \frac{{J_\phi}_\mu {J_\psi}^\mu}{M_{X}^2} \, ,
    \end{equation}
which encode dimension-6 operators involving two Higgs fields, one SM gauge
boson, and a fermion bi-linear current. Diagrammatically, these operators are generated by integrating out the $X$ boson from Feynman diagrams such as that depicted in Fig.~\ref{fig:2Higgs_fermions}.

\begin{figure}[h]
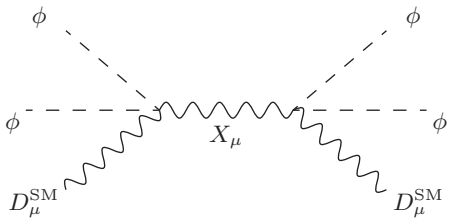

\begin{center}
\begin{axopicture}(200,100)(0,0)
\DashLine(40,80)(75,50){5}
\DashLine(25,50)(75,50){5}
\Photon(40,20)(75,50){3}{5}
\Text(30,85)[c]{$\phi$}
\Text(20,45)[c]{$\phi$}
\Text(28,15)[c]{$D_\mu^{\text{SM}}$}
\Photon(75,50)(125,50){3}{5}
\Text(100,40)[c]{$X_\mu$}
\DashLine(125,50)(160,80){5}
\DashLine(125,50)(175,50){5}
\Photon(125,50)(160,20){3}{5}
\Text(170,85)[c]{$\phi$}
\Text(180,45)[c]{$\phi$}
\Text(172,15)[c]{$D_\mu^{\text{SM}}$}
\end{axopicture}
\end{center}
\caption{$X$-boson mediated process responsible for dimension-6 operators involving four Higgs insertions and either two derivatives or two SM gauge boson insertions.
\label{fig:4Higgs}}    
\end{figure}

Finally, there are terms that are quadratic in the bosonic current $J_\phi^\mu$, 
    \begin{equation}
      {\mathcal L}_\text{SMEFT} \supset 
	- \frac{{J_\phi}_\mu {J_\phi}^\mu}{2M_{X}^2} \, ,
    \end{equation}
which encode dimension-6 operators involving four Higgs fields and two SM covariant derivatives. The corresponding Feynman diagram is given in Fig.~\ref{fig:4Higgs}. 

This accounts, schematically, for the complete set of dimension-6 WCs
  generated by either the \TFHM{} or the \DTFHMp{}\footnote{Note that in
    deriving the WCs we have assumed that the kinetic mixing between the $X$
    boson field strength and the hypercharge field strength
    is negligible at the scale of its derivation, {\em i.e.}\/ at $M_X$.}.
We tabulate all the non-zero WCs generated in this way in Table~\ref{tab:tfhm_wcs} for the
\TFHM{} and in Table~\ref{tab:dtfhmp_wcs} for the \DTFHMp{}.

\begin{table}
\renewcommand{\arraystretch}{1.3}  
   \begin{center}
     \begin{tabular}{|c|c||c|c|} \hline
       WC & value & WC & value \\ \hline
       $C_{ll}^{2222}$&$-\frac{1}{8}$        &
       $ {\bf (C_{lq}^{(1)})}^{22ij}$&$ \frac{1}{12} {\Lambda_{\xi\ ij}^{(d_L)}}$ \\
       $(C_{qq}^{(1)})^{ijkl}$&$ {\Lambda_{\xi\ ij}^{(d_L)}} {\Lambda_{\xi\ kl}^{(d_L)}}\frac{\delta_{ik}\delta_{jl}-2}{72}$ &
       $C_{ee}^{3333}$&$-\frac{1}{2}$ \\     
       $C_{uu}^{3333}$&$-\frac{2}{9}$ &     
       $C_{dd}^{3333}$&$-\frac{1}{18}$ \\     
       $C_{eu}^{3333}$&$\frac{2}{3}$ &     
       $C_{ed}^{3333}$&$-\frac{1}{3}$ \\     
       $(C_{ud}^{(1)})^{3333}$&$\frac{2}{9}$ &
       $C_{le}^{2233}$&$-\frac{1}{2}$ \\     
       $C_{lu}^{2233}$&$\frac{1}{3}$ &     
       $C_{ld}^{2233}$&$-\frac{1}{6}$ \\     
       $C_{qe}^{ij33}$&$\frac{1}{6}{\Lambda_{\xi\ ij}^{(d_L)}}$ &     
       $(C_{qu}^{(1)})^{ij33}$&$-\frac{1}{9} {\Lambda_{\xi\ ij}^{(d_L)}}$ \\     
       $(C_{qd}^{(1)})^{ij33}$&$\frac{1}{18} {\Lambda_{\xi\ ij}^{(d_L)}}$ & 
       $(C_{\phi l}^{(1)})^{22}$&$\frac{1}{4}$ \\
       $(C_{\phi q}^{(1)})^{ij}$&$-\frac{1}{12} {\Lambda_{\xi\ ij}^{(d_L)}}$ &
       $C_{\phi e}^{33}$&$\frac{1}{2}$ \\
       $C_{\phi u}^{33}$&$-\frac{1}{3}$ &
       $C_{\phi d}^{33}$&$\frac{1}{6}$ \\       
       $C_{\phi D}$&$-\frac{1}{2}$ &
       $C_{\phi \Box}$&$-\frac{1}{8}$ \\       
 \hline  \end{tabular}
   \end{center}
     \caption{\label{tab:tfhm_wcs}Non-zero dimension-6 SMEFT WCs predicted by the \TFHM{}, in units of
     $g_{X}^2/M_{X}^2$, in the Warsaw
     basis~\cite{Grzadkowski:2010es}.
     We have highlighted the
     coefficient (for $i=2,\ j=3$) that is primarily responsible for the NCBAs
     in bold font.}
\end{table}
\renewcommand{\arraystretch}{1}

\begin{table}
\renewcommand{\arraystretch}{1.3}  
   \begin{center}
     \begin{tabular}{|c|c||c|c|} \hline
       WC & value & WC & value \\ \hline
       $C_{ll}^{2222}$&$-\frac{25}{72}$        &
       $C_{ll}^{2233}$&$\frac{10}{9}$        \\
       $C_{ll}^{3333}$&$-\frac{8}{9}$        &
       ${\bf (C_{lq}^{(1)})}^{22ij}$&$-\frac{5}{36} {\Lambda_{\xi\ ij}^{(d_L)}}$ \\
       $(C_{lq}^{(1)})^{33ij}$&$\frac{2}{9} {\Lambda_{\xi\ ij}^{(d_L)}}$ &
       $(C_{qq}^{(1)})^{ijkl}$&$ {\Lambda_{\xi\ ij}^{(d_L)}} {\Lambda_{\xi\ kl}^{(d_L)}}\frac{\delta_{ik}\delta_{jl}-2}{72}$ \\
       $C_{ee}^{2222}$&$-\frac{2}{9}$ &
       $C_{ee}^{2233}$&$\frac{10}{9}$ \\
       $C_{ee}^{3333}$&$-\frac{25}{18}$ &
       $C_{uu}^{3333}$&$-\frac{2}{9}$ \\    
       $C_{dd}^{3333}$&$-\frac{1}{18}$ &     
       $C_{eu}^{2233}$&$-\frac{4}{9}$ \\
       $C_{eu}^{3333}$&$\frac{10}{9}$ &            
       $C_{ed}^{2233}$&$\frac{2}{9}$ \\
       $C_{ed}^{3333}$&$-\frac{5}{9}$ &     
       $(C_{ud}^{(1)})^{3333}$&$\frac{2}{9}$ \\
       $C_{le}^{2222}$&$-\frac{5}{9}$ &
       $C_{le}^{2233}$&$\frac{25}{18}$ \\
       $C_{le}^{3333}$&$-\frac{20}{9}$ &
       $C_{le}^{3322}$&$\frac{8}{9}$ \\
       $C_{lu}^{2233}$&$-\frac{5}{9}$ &
       $C_{lu}^{3333}$&$\frac{8}{9}$ \\
       $C_{ld}^{2233}$&$\frac{5}{18}$ &     
       $C_{ld}^{3333}$&$-\frac{4}{9}$ \\     
       $C_{qe}^{ij22}$&$-\frac{1}{9} {\Lambda_{\xi\ ij}^{(d_L)}}$ &
       $C_{qe}^{ij33}$&$\frac{5}{18} {\Lambda_{\xi\ ij}^{(d_L)}}$ \\            
       $(C_{qu}^{(1)})^{ij33}$&$-\frac{1}{9} {\Lambda_{\xi\ ij}^{(d_L)}}$ &
       $(C_{qd}^{(1)})^{ij33}$&$\frac{1}{18} {\Lambda_{\xi\ ij}^{(d_L)}}$ \\
       $(C_{\phi l}^{(1)})^{22}$&$-\frac{5}{12}$ &
       $(C_{\phi l}^{(1)})^{33}$&$\frac{2}{3}$ \\
       $(C_{\phi q}^{(1)})^{ij}$&$-\frac{1}{12} {\Lambda_{\xi\ ij}^{(d_L)}}$ &
       $C_{\phi e}^{22}$&$-\frac{1}{3}$ \\
       $C_{\phi e}^{33}$&$\frac{5}{6}$ &
       $C_{\phi u}^{33}$&$-\frac{1}{3}$ \\
       $C_{\phi d}^{33}$&$\frac{1}{6}$ &      
       $C_{\phi D}$&$-\frac{1}{2}$ \\
       $C_{\phi \Box}$&$-\frac{1}{8}$ & & \\
 \hline  \end{tabular}
   \end{center}
     \caption{\label{tab:dtfhmp_wcs}Non-zero dimension-6 SMEFT WCs predicted by
       the \DTFHMp{},
       in units of
       $g_{X}^2/M_{X}^2$, in the Warsaw basis~\cite{Grzadkowski:2010es}. We have highlighted the
       coefficient that is primarily responsible (for $i=2, j=3$) for the
       NCBAs in bold font. WCs for the original \DTFHM{} may be obtained by
       switching the $l$ indices $2 \leftrightarrow 3$ everywhere.}
   \end{table}
\renewcommand{\arraystretch}{1}  

\section{Global Fits \label{sec:fits}}

\sloppy Given the complete sets of dimension-6 WCs (Tables~\ref{tab:tfhm_wcs} and~\ref{tab:dtfhmp_wcs}) as inputs at the renormalisation scale $M_{X}$,\footnote{Strictly speaking, the
parameters $g_{X}$ and $\theta_{23}$ that we quote are also implicitly {evaluated}
at a renormalisation scale of $M_X$ throughout this paper.} we use the \smelli{} program to calculate
hundreds of
observables and the resulting likelihoods. The \smelli{} program is based upon the observable
calculator
{\tt flavio-2.2.0}~\cite{Straub:2018kue}, using {\tt
  Wilson-2.1}~\cite{Aebischer:2018bkb} for running and matching
WCs using the {\tt WCxf} exchange format~\cite{Aebischer:2017ugx}.

In a particular third family hypercharge  model,
for given values of our {three} input parameters $\theta_{23}$, $g_{X}$,
$M_{X}$, 
the WCs in the tables are converted 
to the non-redundant basis~\cite{Aebischer:2018iyb}
assumed by \smelli{}\footnote{{\tt Jupyter} notebooks
  (from which all data files and plots may be generated) have been
  stored in the {\tt
    anc/} subdirectory of the {\tt arXiv} version of this paper.} (a subset of
the Warsaw basis). 
The renormalisation group equations are then solved in order to run  the WCs
down to the weak scale, at which the EWPOs are calculated.
Most of the EWPOs and correlations are taken from Ref.~\cite{ALEPH:2005ab}
by \smelli{}, which neglects the relatively small theoretical uncertainties in
EWPOs.
The EWPOs have not been averaged over lepton flavour, since lepton flavour
non-universality is a key feature of any model of the NCBAs, including those
built on third family hypercharge which we consider.

{We have updated the data used by {\tt flavio-2.2.0} with 2021 LHCb
  measurements of $BR(B_{d,s} \rightarrow \mu^+ \mu^-)$
  taken on 9 fb$^{-1}$ of LHC Run II data~\cite{LHCbtalk} 
  by using the two dimensional Gaussian fit to current CMS,
  ATLAS and LHCb measurements presented in
  Ref.~\cite{Altmannshofer:2021qrr}:
  \begin{eqnarray}
  BR(B_s \rightarrow \mu^+ \mu^-) &=& (2.93 \pm 0.35) \times 10^{-9},
  \nonumber \\
  BR(B^0 \rightarrow \mu^+ \mu^-) &=& (0.56 \pm 0.70) \times 10^{-10},
  \end{eqnarray}
  with an error correlation of $\rho = -0.27$.
  The most recent
  measurement by LHCb in the di-lepton invariant mass squared bin
  $1.1<Q^2/\text{GeV}^2 < 6.0$ is
  \begin{equation}
  R_K= 0.846^{+0.042}_{-0.039}{}^{+0.013}_{-0.012},
  \end{equation} where the first error is statistical and the second
  systematic~\cite{Aaij:2021vac} (this measurement alone has a 3.1$\sigma$ tension
  with the SM prediction of 1.00). We incorporate this new measurement
  by
  fitting the log likelihood function presented in Ref.~\cite{Aaij:2021vac}
  with a quartic polynomial.}

The SMEFT weak scale WCs are then matched to the
weak effective theory and renormalised down to the scale of bottom mesons
using QCD$\times$QED\@. Observables relevant to the NCBAs are calculated at this scale.
\smelli{} then organises the calculation of the $\chi^2$ statistic to quantify
a distance (squared) between the theoretical prediction and experimental observables in
units of the uncertainty.
In calculating the $\chi^2$ value,
experimental correlations between different observables are parameterised and
taken into account. 
Theoretical uncertainties are modelled as being multi-variate Gaussians; they include the effects
of varying nuisance parameters and are approximated to be independent of new
physics. Theory uncertainties and experimental uncertainties are then
combined in quadrature.

\begin{figure}
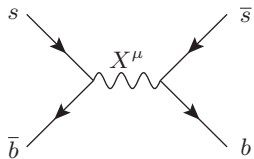

\begin{center}
  \begin{axopicture}(80,55)(-5,-5)
    \Line[arrow](0,50)(25,25)
    \Line[arrow](25,25)(0,0)
    \Line[arrow](75,50)(50,25)
    \Line[arrow](50,25)(75,0)
    \Photon(25,25)(50,25){3}{3}
    \Text(37.5,33)[c]{$X^\mu$}
    \Text(-5,0)[c]{$\overline{b}$}
    \Text(-5,50)[c]{$s$}
    \Text(82,50)[c]{$\overline{s}$}
    \Text(82,0)[c]{$b$}    
  \end{axopicture}
\caption{\label{fig:bsbsbar} An $X$-boson mediated
  contribution to $B_s - \overline{B_s}$ mixing.}
\end{center}
\end{figure}
We note that an important constraint on $Z^\prime$ models that explain the
NCBAs is that from $\Delta m_s$ (included by \smelli{} in the category of
`quarks' observables), deriving from measurements of $B_s-\overline{B_s}$
mixing, because of the {tree-level BSM contribution to the} process depicted in Fig.~\ref{fig:bsbsbar}.
The impact of this constraint has significantly varied over the last decade,
to a large degree because of numerically rather different lattice or theory
inputs used to extract 
the measurement~\cite{Bazavov:2016nty,DiLuzio:2017fdq,King:2019lal}.
Here, we are wedded to the calculation and inputs
used by \smelli{}, allowing some tension in $\Delta m_s$ to be traded against
tension present in the NCBAs.

As we shall see, in all the models that we consider the global fit is fairly insensitive to
$M_{X}$, provided we specify $M_{X} > 2$ TeV or so in order to be sure
to not contravene
ATLAS di-muon searches~\cite{Aad:2019fac,Allanach:2019iiy,Allanach:2019mfl}. We will demonstrate this
insensitivity to $M_X$ below (see Figs.~\ref{fig:tfhm_mzp} and~\ref{fig:dtfhmp_mzp}), but for now we 
shall pick $M_{X}=3$ TeV and scan 
over the pair $(g_{X} \times \text{3
  TeV}/M_{X})$ and $\theta_{23}$. Since the WCs at $M_{X}$ all scale like $g_{X}/M_{X}$, the results will approximately hold at different values of $M_{X}$ provided that $g_{X}$ is scaled linearly with $M_{X}$. The running between $M_{X}$ and the weak scale breaks this scaling, but such effects derive from loop corrections $\propto (1/16 \pi^2)\ln (M_{X}/M_Z)$ and are thus negligible to a good approximation.

\subsection{\TFHM{} fit results\label{sec:tfhm_results}}

\begin{table}
  \begin{center}
    \begin{tabular}{|c|ccc|} \hline
    data set & $\chi^2$ & $n$ & $p-$value \\ \hline
    \input{anc/tfhm_fit.tex} \hline
  \end{tabular}
  \caption{\label{tab:tfhm_fit} Goodness of fit for the different data sets
    we consider for the \TFHM{} as calculated by \smelli{} for $M_{X}=3$
    TeV.
    We display the total 
    $\chi^2$ for each data set along with the number of observables $n$ and
    the data set's $p-$value.  The data set named `quarks' contains $BR(B_s
    \rightarrow \phi \mu^+ \mu^-)$, 
    $BR(B_s\rightarrow \mu^+\mu^-)$,  $\Delta m_s$ and various differential
    distributions in $B\rightarrow K^{(\ast)} \mu^+\mu^-$ decays among others, whereas `LFU
    FCNCs' contains $R_{K^{(\ast)}}$ and some $B$ meson decay branching ratios
  into di-taus. Our data sets are identical to those defined by \smelli{}
  and we refer the curious reader to its manual~\cite{Aebischer:2018iyb},
  where the observables are enumerated.     We have updated to $R_K$ and $BR(B_{s,d}\rightarrow \mu^+ \mu^-)$ with the
    latest LHCb measurements as detailed in \S\ref{sec:fits}.
Two free parameters of the model were
    fitted: $\theta_{23}=\TFHMtheta{}$ and $g_{X}=\TFHMgzp{}$.}
  \end{center}
  \end{table}
The result of fitting $\theta_{23}$ and $g_{X}$ for $M_{X}=3$
TeV is shown in Table~\ref{tab:tfhm_fit} for the \TFHM{}. The `global' $p-$value is
calculated by assuming a $\chi^2$ distribution with $n-2$ degrees of freedom,
since two parameters were optimised.
The fit is
encouragingly of a much better quality than the one of the SM\@. We see that the
fits to the EWPOs and NCBAs are simultaneously reasonable.

\begin{figure}
  \begin{center}
    \unitlength=\columnwidth
    \begin{picture}(0.8, 1.4)(0,0)
      \put(0,0){\includegraphics[width=0.8 \columnwidth]{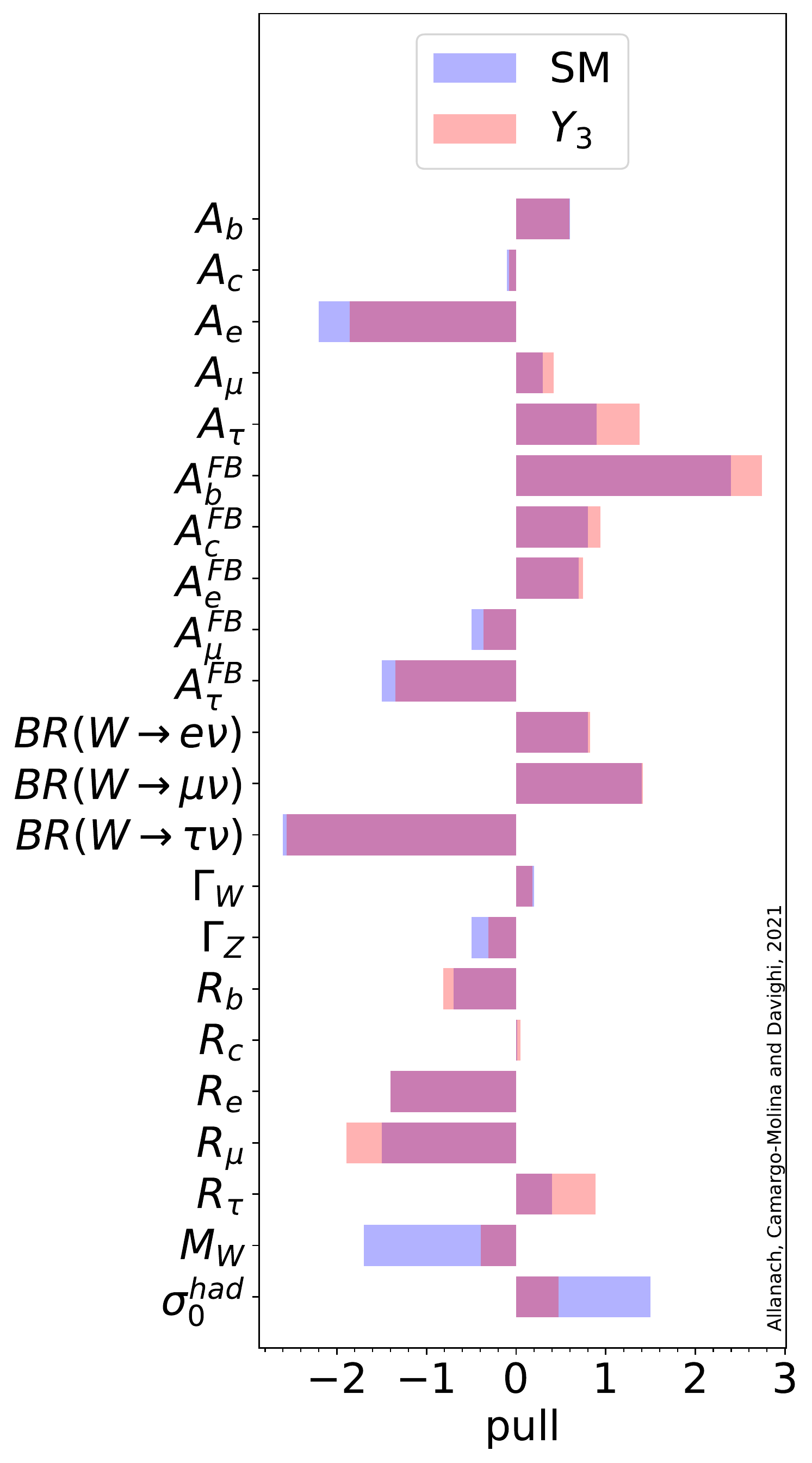}}
      \end{picture}
  \end{center}
  \caption{Pulls in the EWPOs for the \TFHM{} $M_{X}=3$ TeV best-fit point:
     $\theta_{23}=\TFHMtheta{}$, $g_{X}=\TFHMgzp{}$.
    The pull is defined to be the theory prediction minus the central
    value of the observation, divided by the combined theoretical and
    experimental uncertainty, neglecting any
    correlations with other observables.
    \label{fig:tfhm_ewpo}}
  \end{figure}
The EWPOs are shown in more detail in Fig.~\ref{fig:tfhm_ewpo}, in which we
compare some
pulls in the SM fit versus
the \TFHM{} best-fit point. 
We see that there is some improvement in the prediction of the $W$-boson
mass, which the SM fit predicts is almost 2$\sigma$ too low (as manifest in the $\rho$-parameter being measured to be slightly larger than one~\cite{Zyla:2020zbs}, {for $M_Z$ taken to be fixed to its SM value}).
The easing of this tension in $M_W$ is due precisely to the $Z-Z^\prime$
mixing in the (D)\TFHM{}s. The non-zero value of the SMEFT coefficient
$C_{\phi D}$ breaks custodial symmetry, resulting in a shift of the
$\rho$-parameter away from its tree-level SM value of one, to~\cite{Davighi:2020nhv} 
\begin{equation}
(\rho_0)_{Y_3} =  1 -
C_{\phi D} v^2 / 2  =1 + v^2 g_X^2/(4m^2_X).
\end{equation}
where $v$ is the SM Higgs VEV\@.
Rather than being dangerous, as might reasonably have been guessed, it turns
out that this BSM contribution to $\rho_0$ is in large part responsible for
the \TFHM{} fitting the EWPOs approximately as well as the SM does.

We also see that
$\sigma_0^{had}$, the $e^+ e^-$ scattering cross-section to hadrons at a
centre-of-mass energy of $M_Z$, is better fit by the \TFHM{} than the SM\@. 
Although the other EWPOs have some small deviations from their SM fits, the
overall picture is that the \TFHM{} best-fit point has an electroweak fit
similar to that of the SM. 

\begin{figure}
  \begin{center}
    \unitlength=\columnwidth
    \begin{picture}(1,0.67)(0,0)
      \put(0,0){\includegraphics[width=\columnwidth]{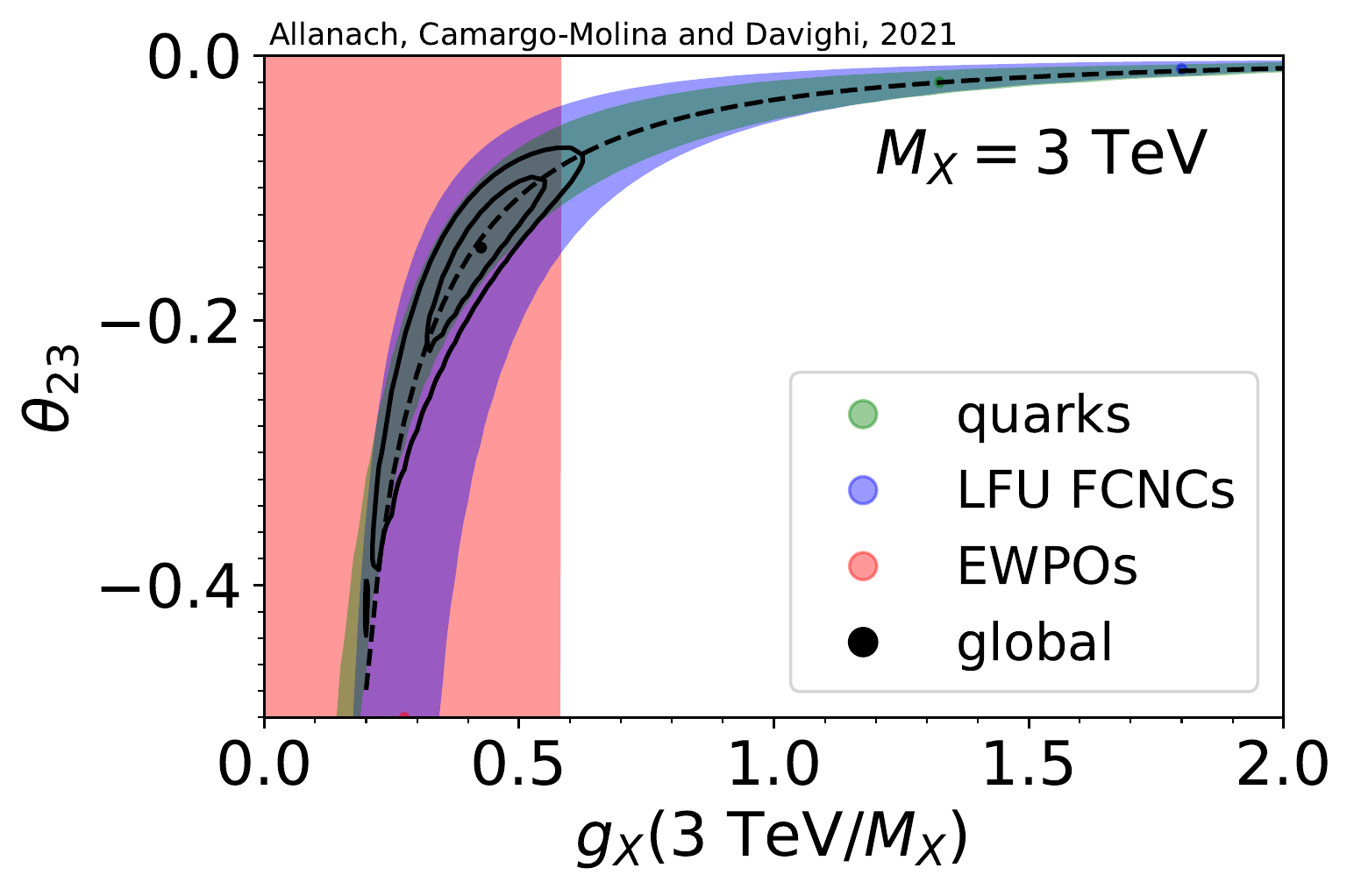}}
      \end{picture}
  \end{center}
  \caption{Two parameter fit to the \TFHM{} for $M_{X}=3$ TeV. Shaded
    regions are those preferred by the data set in the legend at the 95$\%$
    confidence level (CL).
    The global fit is shown by the solid curves, where  the  
    inner (outer) curves 
    show the 70$\%$(95$\%$) CL regions, respectively.
    The set named `quarks' contains $BR(B_s
    \rightarrow \phi \mu^+ \mu^-)$, 
    $BR(B_s\rightarrow \mu^+\mu^-)$,  $\Delta m_s$ and various differential
    distributions in $B\rightarrow K^{(\ast)} \mu^+\mu^-$ decays among others, whereas `LFU
    FCNCs' contains $R_{K^{(\ast)}}$ and some $B$ meson decay branching ratios
    into di-taus. Our sets are identical to those defined by \smelli{}
    and we refer the curious reader to its manual~\cite{Aebischer:2018iyb},
    where the observables are enumerated.
    We have updated $R_K$ and $BR(B_{s,d}\rightarrow \mu^+ \mu^-)$ with the
    latest LHCb measurements as detailed in \S\ref{sec:fits}.
    The black dot marks the locus of the best-fit point.
    \label{fig:tfhm_global}}
  \end{figure}

In order to see which areas of parameter space are favoured by the different
sets of constraints, we provide Fig.~\ref{fig:tfhm_global}.
The figure shows that the EWPOs and different sets of NCBA data all overlap at
the 95$\%$ CL\@. The best-fit point has a total $\chi^2$ of 43 less than {that} of the
SM and is marked by a black dot. {The separate data set contributions to $\chi^2$ at} this point are listed in
Table~\ref{tab:tfhm_fit}. In order to calculate 70$\%$ (95$\%$) CL bounds in the
2-dimensional parameter plane, we
draw contours of
$\chi^2$ equal to the best-fit value plus
$2.41$ {($5.99$)} respectively, {using the combined $\chi^2$ incorporating all the datasets}. 

\begin{figure}
  \begin{center}
    \unitlength=\columnwidth
    \begin{picture}(1,0.5)(0,0)
      \put(0,0){\includegraphics[width=\columnwidth]{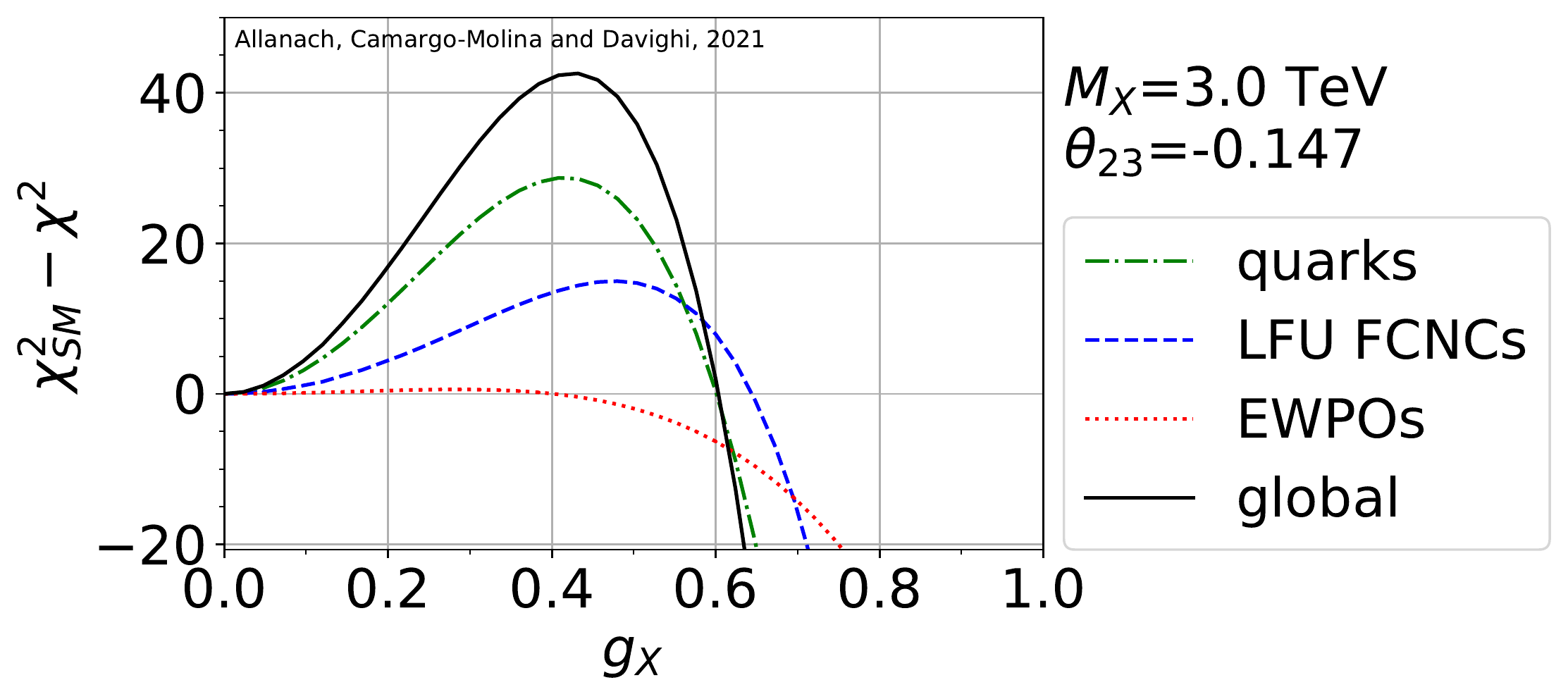}}
      \end{picture}
  \end{center}
  \caption{\TFHM{} $\Delta \chi^2$ contributions in the vicinity of the best-fit
    point as a function of $g_{X}$.
    \label{fig:tfhm_gzp}}
  \end{figure}
\begin{figure}
  \begin{center}
    \unitlength=\columnwidth
    \begin{picture}(1,0.5)(0,0)
      \put(0,0){\includegraphics[width=\columnwidth]{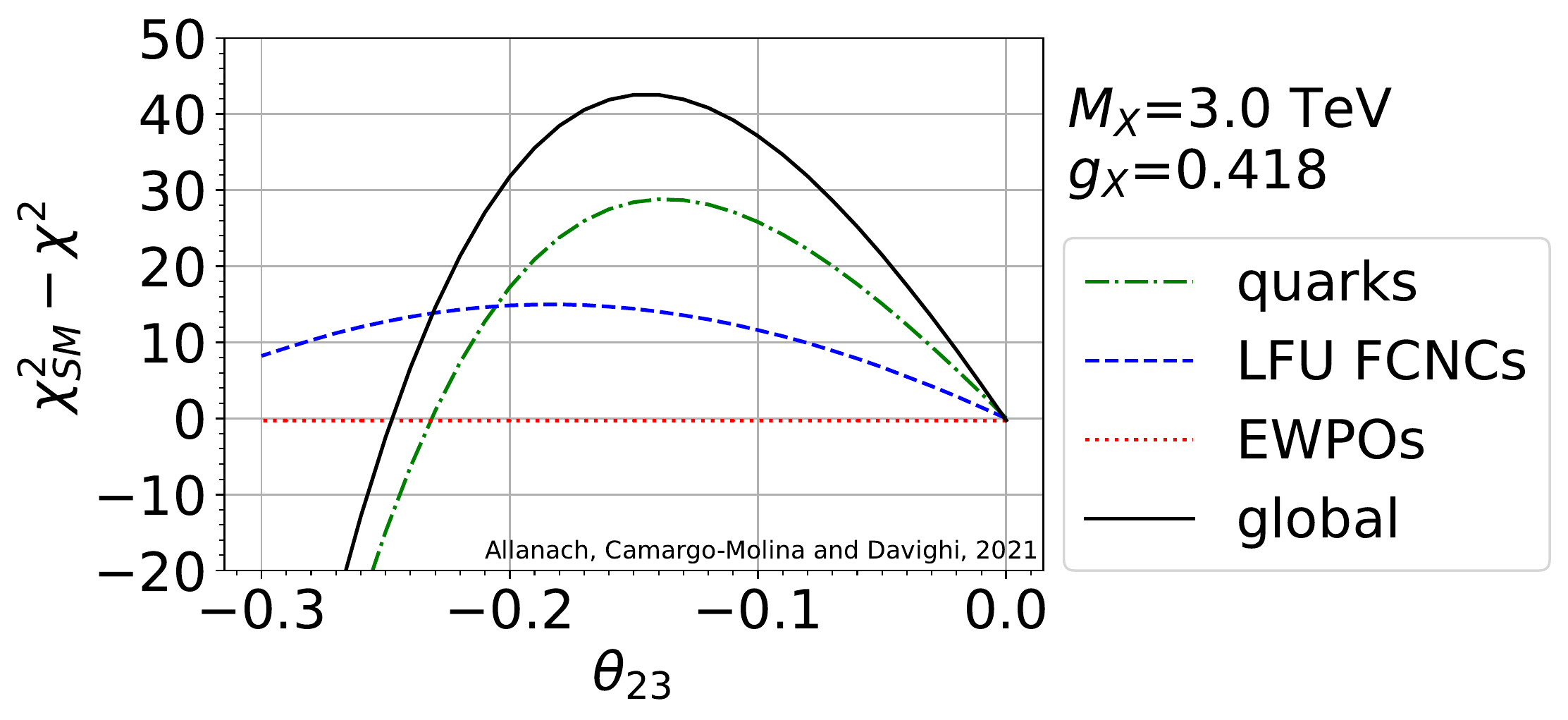}}
      \end{picture}
  \end{center}
  \caption{\TFHM{} $\Delta \chi^2$ contributions in the vicinity of the best-fit
    point as a function of $\theta_{23}$.
    \label{fig:tfhm_theta}}
  \end{figure}
\begin{figure}
  \begin{center}
    \unitlength=\columnwidth
    \begin{picture}(1,0.5)(0,0)
      \put(0,0){\includegraphics[width=\columnwidth]{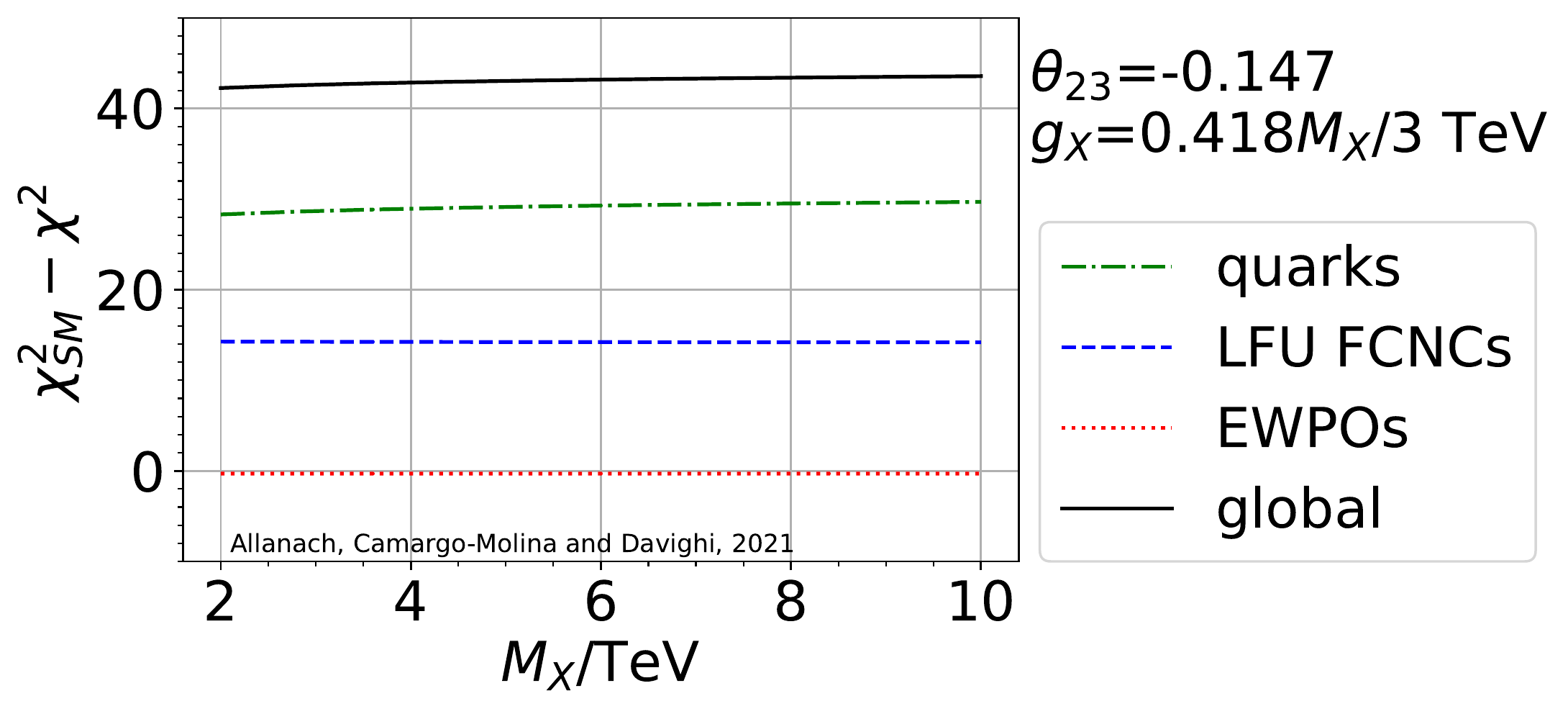}}
      \end{picture}
  \end{center}
  \caption{\TFHM{} $\Delta \chi^2$ contributions in the vicinity of the best-fit
    point as a function of $M_{X}$, where $g_{X}$ has been
    scaled linearly. 
    \label{fig:tfhm_mzp}}
  \end{figure}
We {further} study the  different $\chi^2$ contributions for the \TFHM{} in the
vicinity of
the best-fit point in
Figs.~\ref{fig:tfhm_gzp}-\ref{fig:tfhm_mzp}. {From} Fig.~\ref{fig:tfhm_gzp}, we see that {large couplings}
$g_{X}>0.6$ are disfavoured by EWPOs as well as the NCBAs. {From}
Fig.~\ref{fig:tfhm_theta} we see that the EWPOs are insensitive to the
value of $\theta_{23}$ in the vicinity of the best-fit point but the NCBAs are
not.
At large $-\theta_{23}$ the \TFHM{} suffers {due to} a bad fit to
the $B_s-\overline{B}_s$ mixing observable $\Delta m_s$.
In Fig.~\ref{fig:tfhm_mzp}, we demonstrate the approximate insensitivity of $\chi^2$
near the best-fit point to $M_{X}$, provided that $g_{X}$ is
scaled linearly with it. 

\begin{figure}
  \begin{center}
    \unitlength=\columnwidth
    \begin{picture}(1, 0.7)(0,0)
      \put(0,0){\includegraphics[width=\columnwidth]{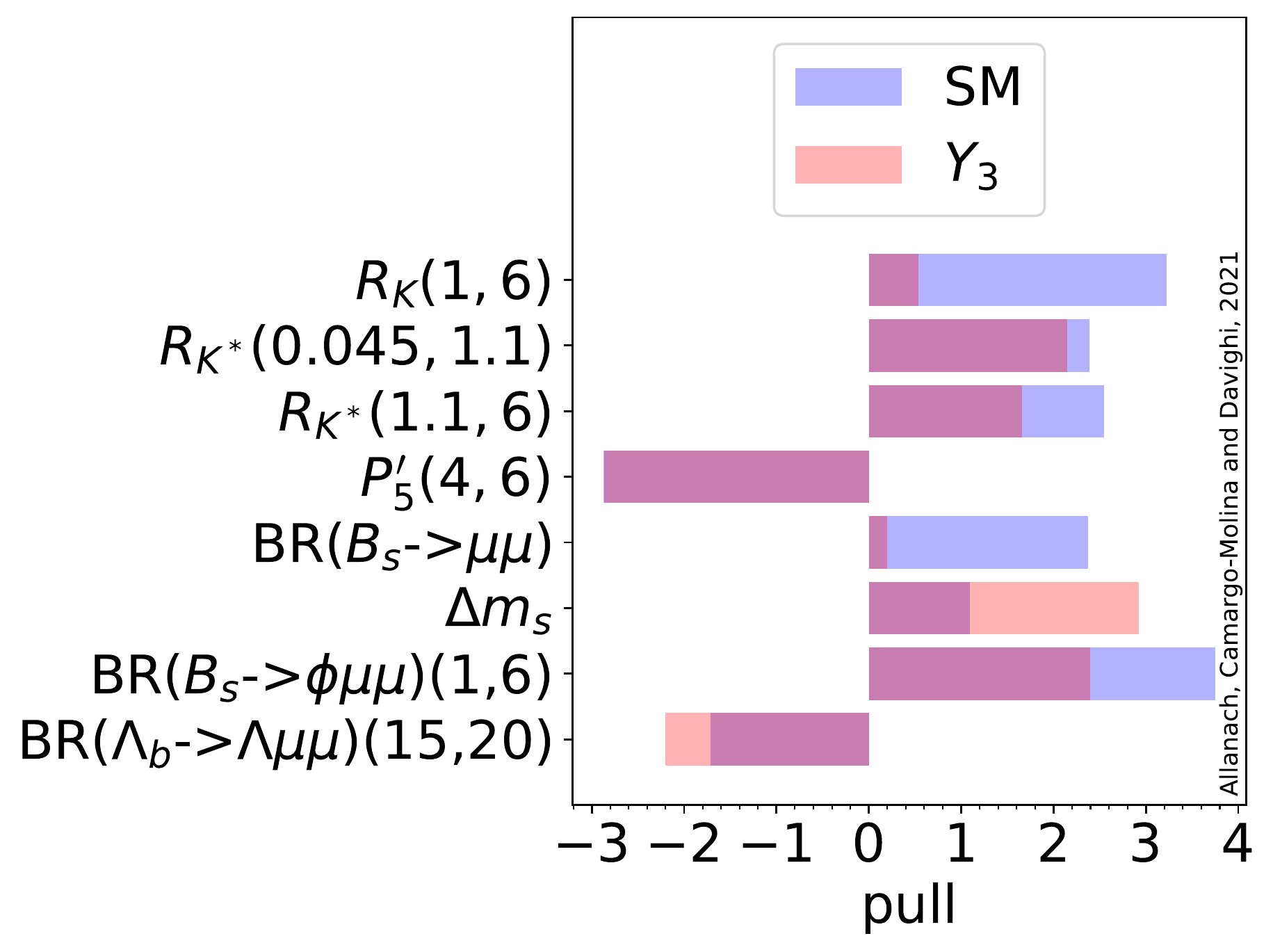}}
      \end{picture}
  \end{center}
  \caption{Pulls of interest for the \TFHM{} $M_{X}=3$ TeV best-fit point:
     $\theta_{23}=\TFHMtheta{}$,
    $g_{X}=\TFHMgzp{}$.
    The pull is defined to be the theory prediction minus the central
    value of the observation, divided by the combined theoretical and
    experimental uncertainty, neglecting any
    correlations with other observables. Numbers in brackets after the
    observable name refer to minimum and maximum values of
    $m_{\mu\mu}^2$
    of the bin in GeV$^2$, respectively (many other bins and observables are
    also used to compute the global likelihood). 
    \label{fig:tfhm_int}}
  \end{figure}

{Finally,} we display some individual observables of interest in
Fig.~\ref{fig:tfhm_int} at the \TFHM{} best-fit point.
While some of the prominent NCBA measurements (for example
$R_K$ in the bin of {$m_{\mu\mu}^2$} between 1.1 GeV$^{2}$ and 6 GeV$^{2}$) fit
considerably better than the SM, we see that this is partly compensated by
a worse fit in $\Delta m_s$, {as is the case for many $Z^\prime$ models for
  the NCBAs.} {The $P_5^\prime$ observable (derived in terms of angular
distributions of $B^0\rightarrow K^\ast \mu^+\mu^-$ decays~\cite{Matias:2012xw,Descotes-Genon:2015uva}) shows no
significant change from the SM prediction in the bin that deviates the most
significantly from experiment: $m_{\mu\mu}^2 \in (4,\ 6)$ GeV$^2$, as measured by
LHCb~\cite{Aaij:2015oid} and ATLAS~\cite{Aaboud:2018krd}.
The fit to $BR(\Lambda_b
\rightarrow \Lambda \mu^+ \mu^-)$ is slightly 
worse than that of the SM in one particular bin, as shown in the figure.
Some other flavour observables in the flavour sector, notably various bins of $BR(B\rightarrow
K^{(\ast)}\mu^+ \mu^-)$, show some small differences in pulls between the SM
and the \TFHM{}. Whilst there are many of these and in aggregate they make a
difference to the overall $\chi^2$, there is no small set of observables that
provide the driving force and
so we neglect to show them\footnote{The interested reader can find values for all
observables and pulls in the {\tt Jupyter} notebook in the ancillary
information associated with the {\tt arXiv} version of this paper.}. We shall
now turn to the \DTFHMp{} fit results, where these comments about flavour observables
also apply.
}

\subsection{\DTFHMp{} fit results\label{sec:dtfhmp_results}}

\begin{table}
  \begin{center}
    \begin{tabular}{|c|ccc|} \hline
    data set & $\chi^2$ & $n$ & $p-$value \\ \hline
    \input{anc/dtfhmp_fit.tex} \hline
  \end{tabular}
  \caption{\label{tab:dtfhmp_fit} Goodness of fit for the different data sets
    we consider for the \DTFHMp{}, as calculated by \smelli{} for
    $M_{X}=3$ TeV. We display the total  
    $\chi^2$ for each data set along with the number of observables $n$ and
    the data set's $p-$value. The set named `quarks' contains $BR(B_s
    \rightarrow \phi \mu^+ \mu^-)$, 
    $BR(B_s\rightarrow \mu^+\mu^-)$,  $\Delta m_s$ and various differential
    distributions in $B\rightarrow K^{(\ast)} \mu^+\mu^-$ decays among others, whereas `LFU
    FCNCs' contains $R_{K^{(\ast)}}$ and some $B$ meson decay branching ratios
  into di-taus. {Our sets are identical to those defined by \smelli{}
  and we refer the curious reader to its manual~\cite{Aebischer:2018iyb},
  where the observables are enumerated.}
    We have updated $R_K$ and $BR(B_{s,d}\rightarrow \mu^+ \mu^-)$ with the
    latest LHCb measurements as detailed in \S\ref{sec:fits}.
    Two free parameters of the model were
    fitted: $\theta_{23}=\DTFHMptheta{}$ and $g_{X}=\DTFHMpgzp{}$.} 
  \end{center}
  \end{table}
We {summarise the quality of the fit for the \DTFHMp{} at the best-fit point, for $M_{X}=3$ TeV,} in
Table~\ref{tab:dtfhmp_fit}. We see a much improved fit as compared to the SM (by a
$\Delta \chi^2=39$) and a similar (but slightly worse) quality of fit 
compared to the \TFHM{}, as a
comparison with Table~\ref{tab:tfhm_fit} shows. 

\begin{figure}
  \begin{center}
    \unitlength=\columnwidth
    \begin{picture}(1,0.75)(0,0)
      \put(0,0){\includegraphics[width=\columnwidth]{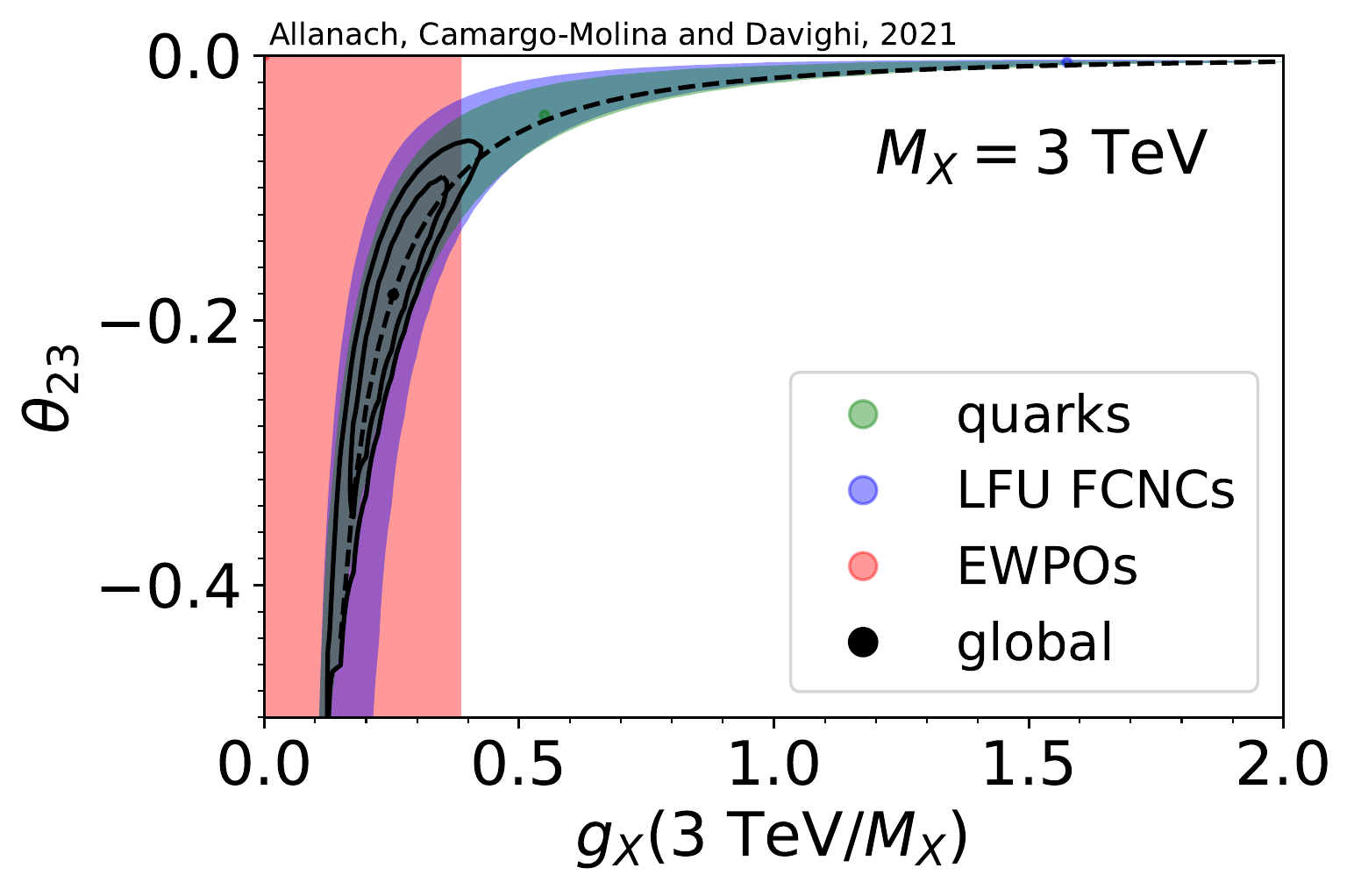}}
      \end{picture}
  \end{center}
  \caption{Two parameter fit to the \DTFHMp{}  for $M_{X}=3$ TeV. Shaded
    regions are those preferred by the data set in the legend at the 95$\%$
    confidence level (CL).
    The global fit is shown by the solid curves, where    the
    inner (outer) curves 
    show the 70$\%$ (95$\%$) CL regions, respectively.
    The set named `quarks' contains $BR(B_s
    \rightarrow \phi \mu^+ \mu^-)$, 
    $BR(B_s\rightarrow \mu^+\mu^-)$,  $\Delta m_s$ and various differential
    distributions in $B\rightarrow K^{(\ast)} \mu^+\mu^-$ decays among others, whereas `LFU
    FCNCs' contains $R_{K^{(\ast)}}$ and some $B$ meson decay branching ratios
    into di-taus. Our sets are identical to those defined by \smelli{}
    and we refer the curious reader to its manual~\cite{Aebischer:2018iyb},
    where the observables are enumerated. We have updated $R_K$ and $BR(B_{s,d}\rightarrow \mu^+ \mu^-)$ with the
    latest LHCb measurements as detailed in \S\ref{sec:fits}.
    The black dot marks the locus of the best-fit point.
    \label{fig:dtfhmp_global}}
  \end{figure}
The constraints upon the parameters $\theta_{23}$ and $g_{X}$ are shown
in Fig.~\ref{fig:dtfhmp_global}. Although the figure {is} for
$M_{X}=3$ TeV, {the picture} remains approximately the same for
$2 < M_{X}/\text{TeV} < 10$. We see that, as is the case for the
\TFHM{}, there is a region of overlap of the 95$\%$ CL regions of all of the
constraints. 

\begin{figure}
  \begin{center}
    \unitlength=\columnwidth
    \begin{picture}(0.8, 1.4)(0,0)
      \put(0,0){\includegraphics[width=0.8 \columnwidth]{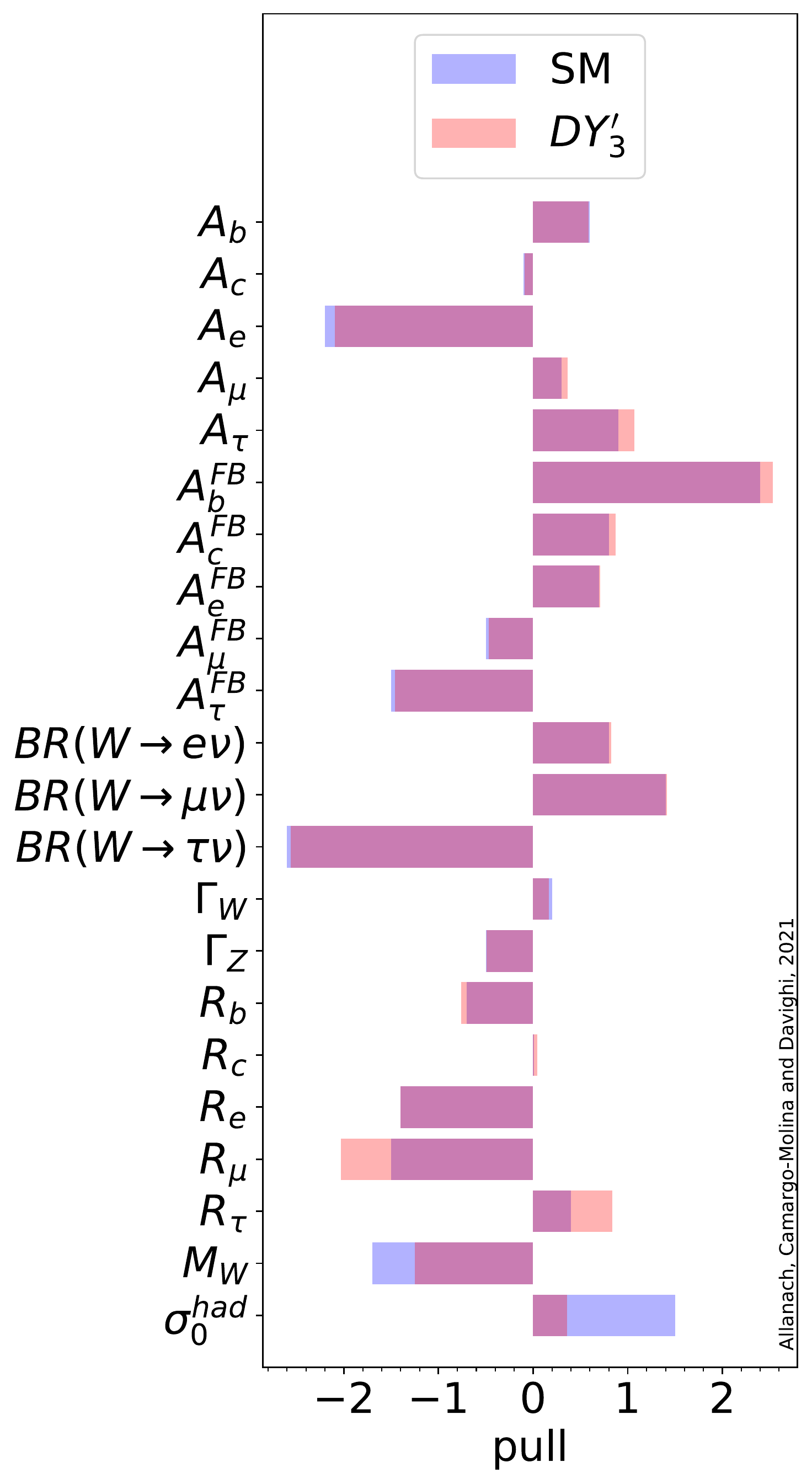}}
      \end{picture}
  \end{center}
  \caption{Pulls in the EWPOs for the $M_{X}=3$ TeV \DTFHMp{} 
     best-fit point:
    $g_{X}=\DTFHMpgzp{}$, $\theta_{23}=\DTFHMptheta{}$.
    The pull is defined to be the theory prediction minus the central
    value of the observation, divided by the combined theoretical and
    experimental uncertainty, neglecting any
    correlations with other observables.
    \label{fig:dtfhmp_best_ewpo}}
  \end{figure}
The pulls in the EWPOs for the best-fit point of the \DTFHMp{} {are} shown in
Fig.~\ref{fig:dtfhmp_best_ewpo}. Like the \TFHM{} above, we see a fit
comparable in quality to that of the SM\@. Again, the \DTFHMp{} predicts $M_W$  to be
a little higher than in the SM, agreeing slightly better with the experimental measurement.

\begin{figure}
  \begin{center}
    \unitlength=\columnwidth
    \begin{picture}(1,0.5)(0,0)
      \put(0,0){\includegraphics[width=\columnwidth]{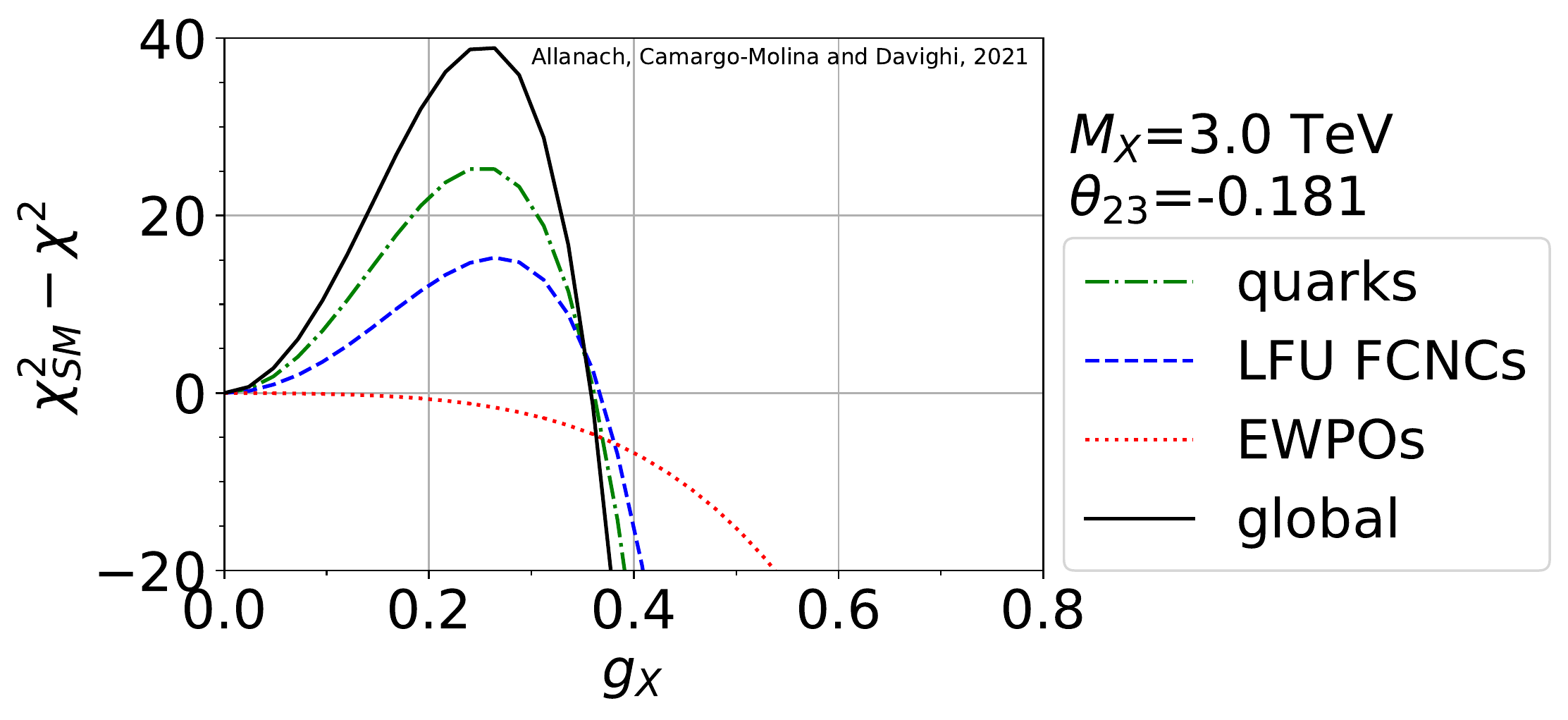}}
      \end{picture}
  \end{center}
  \caption{$\Delta \chi^2$ contributions in the vicinity of the  \DTFHMp{} best-fit
    point as a function of $g_{X}$. 
    \label{fig:dtfhmp_gzp}}
  \end{figure}
\begin{figure}
  \begin{center}
    \unitlength=\columnwidth
    \begin{picture}(1,0.5)(0,0)
      \put(0,0){\includegraphics[width=\columnwidth]{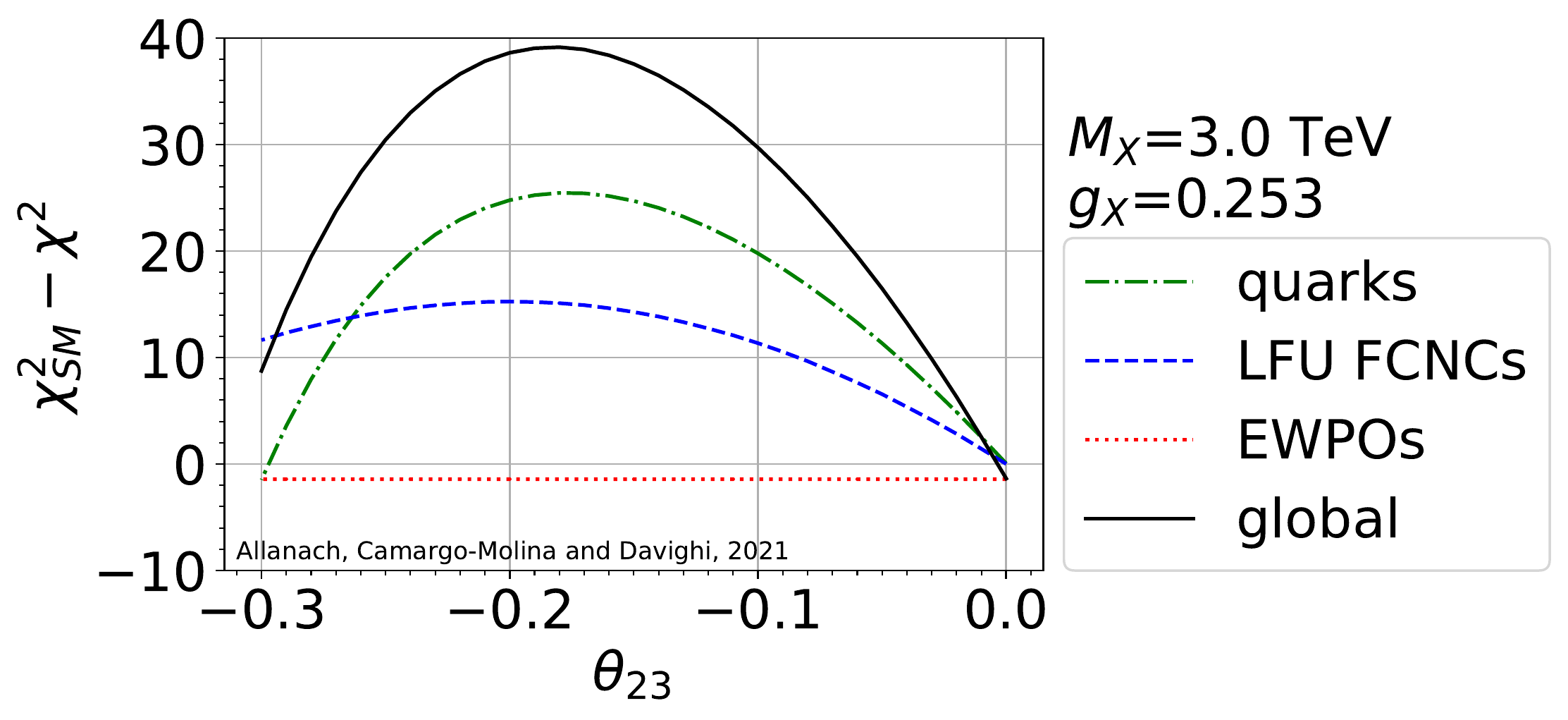}}
      \end{picture}
  \end{center}
  \caption{$\Delta \chi^2$ contributions in the vicinity of the  \DTFHMp{} best-fit
    point as a function of $\theta_{23}$. 
    \label{fig:dtfhmp_theta}}
  \end{figure}
\begin{figure}
  \begin{center}
    \unitlength=\columnwidth
    \begin{picture}(1,0.5)(0,0)
      \put(0,0){\includegraphics[width=\columnwidth]{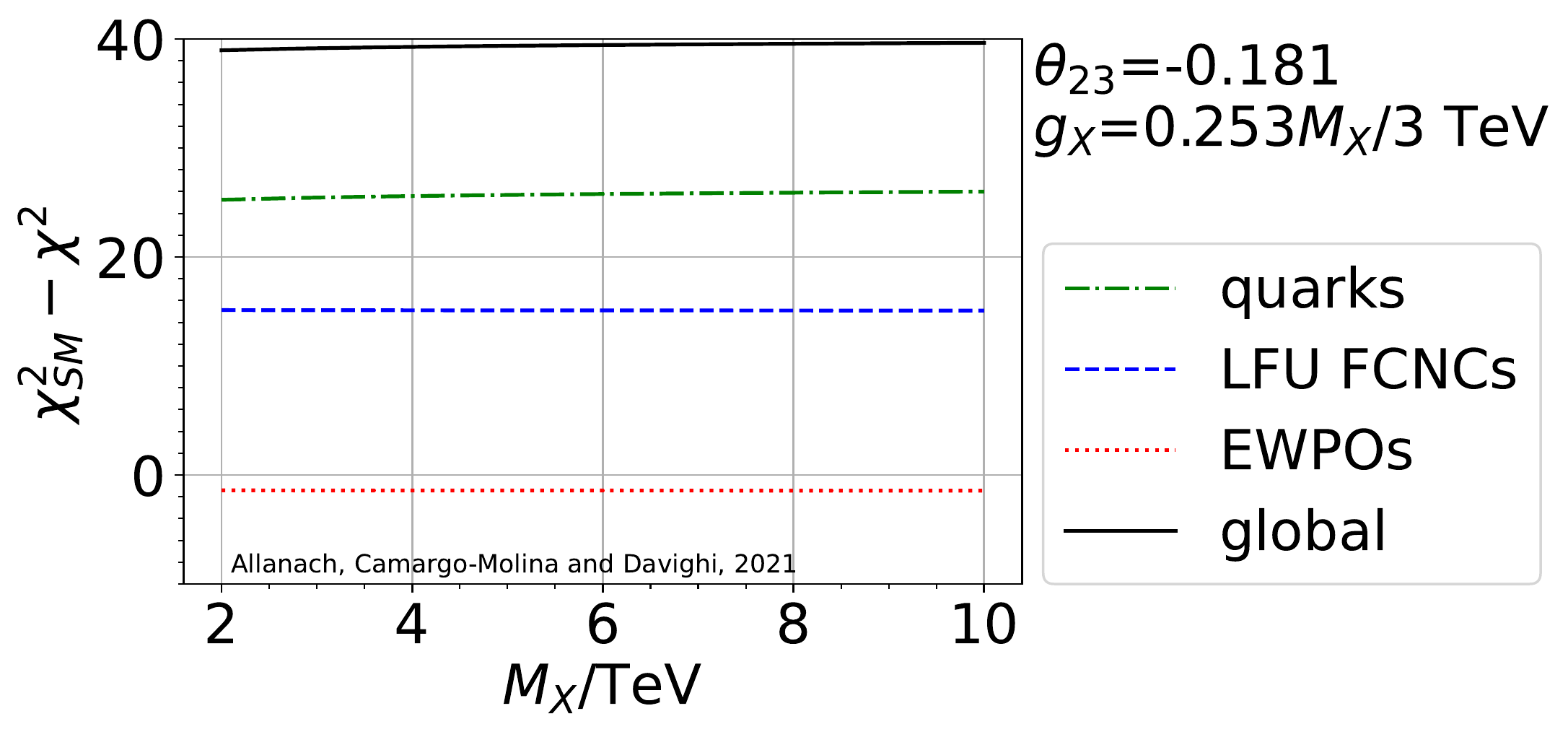}}
      \end{picture}
  \end{center}
  \caption{$\Delta \chi^2$ contributions in the vicinity of the  \DTFHMp{} best-fit
    point as a function of $M_{X}$, where $g_{X}$ has been
    scaled linearly. 
    \label{fig:dtfhmp_mzp}}
  \end{figure}
The behaviour of the fit in various directions in parameter space around the
best-fit point is shown in Figs.~\ref{fig:dtfhmp_gzp}-\ref{fig:dtfhmp_mzp}.
Qualitatively, this behaviour is similar to that of the \TFHM{}: the
EWPOs and NCBAs imply that $g_{X}$ should not become too large. {The mixing observable} $\Delta
m_s$ prevents $-\theta_{23}$ from becoming too large, and the fits are insensitive
to $M_{X}$ varied between 2 TeV and 10 TeV so long as $g_{X}$ is
scaled linearly with $M_{X}$. 

\begin{figure}
  \begin{center}
    \unitlength=\columnwidth
    \begin{picture}(1, 0.75)(0,0)
      \put(0,0){\includegraphics[width=\columnwidth]{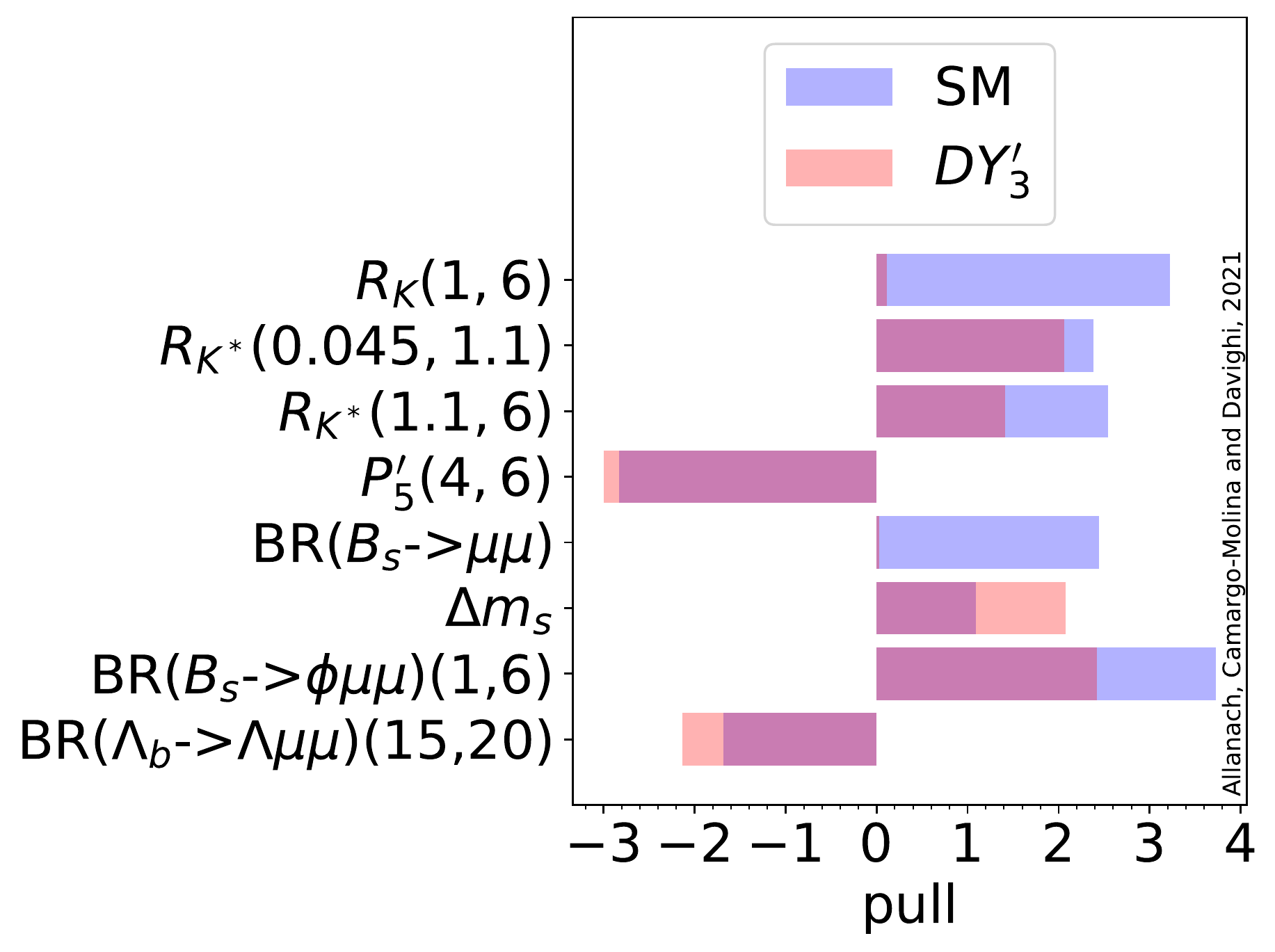}}
      \end{picture}
  \end{center}
  \caption{Pulls of interest for the $M_{X}=3$ TeV \DTFHMp{} best-fit point:
    $g_{X}=\DTFHMpgzp{}$,
    $\theta_{23}=\DTFHMptheta{}$.
    The pull is defined to be the theory prediction minus the central
    value of the observation, divided by the combined theoretical and
    experimental uncertainty, neglecting any
    correlations with other observables. Numbers in brackets after the
    observable name refer to minimum and maximum values of
    {$m_{\mu\mu}^2$} in GeV$^2$, respectively (many other bins and observables are
    also used to compute the global likelihood). 
    \label{fig:dtfhmp_int}}
  \end{figure}
Fig.~\ref{fig:dtfhmp_int} shows various pulls of interest at the best-fit
point of the \DTFHMp{}. Better fits (than the SM) to several NBCA observables
are partially counteracted by a worse fit to the $\Delta m_s$ observable.

\subsection{Original \DTFHM{} fit results}\label{sec:DTFHMresults}
We display the overall fit quality of the original \DTFHM{} in Table~\ref{tab:dtfhm_fit}.
By  comparison with Tables~\ref{tab:sm_fit} {and}~\ref{tab:tfhm_fit}
we see that although its predictions still fit the data significantly better than
the SM ($\Delta \chi^2$ is 32), the original \DTFHM{} does not achieve as good
fits as the other models. For the sake of brevity, we have refrained from
including plots for it\footnote{These may be found within a {\tt
    Jupyter} notebook in the {\tt anc}
  sub-directory of the {\tt arXiv} submission of this paper.}. {Instead, it is more enlightening to understand the reason behind this slightly
worse fit, which is roughly as follows.}
The coupling of {the} $X$ boson to muons in the original \DTFHM{} is close to
vector-like, {\em viz.}\ ${\mathcal L} = 
g_X/6 (\overline{\mu} \slashed{X} (5P_L + 4P_R)) \mu + \ldots$ (where $P_L,P_R$ are left-handed and right-handed projection operators, respectively),
which is {slightly} less preferred by the \smelli{} fits than an $X$ boson coupled more
strongly to left-handed muons~\cite{Aebischer:2018iyb}. {This preference is in large part due to
the experimentally measured value of $BR(B_s\rightarrow \mu^+
\mu^-)$, which is somewhat lower than the SM prediction~\cite{Aaboud:2018mst,Chatrchyan:2013bka,CMS:2014xfa,Aaij:2017vad}, and is sensitive only to the axial component of the coupling to muons.}
Compared to the \TFHM{} and the \DTFHMp{}, the fit to $BR(B_s\rightarrow \mu^+
\mu^-)$ is worse when the \DTFHM{} fits other observables well.
\begin{table}
  \begin{center}
    \begin{tabular}{|c|ccc|} \hline
    data set & $\chi^2$ & $n$ & $p-$value \\ \hline
    \input{anc/dtfhm_fit.tex} \hline
  \end{tabular}
  \caption{\label{tab:dtfhm_fit} Goodness of fit for the different data sets
    we consider for the {original} \DTFHM{}, as calculated by \smelli{}  for $M_{X}=3$
    TeV. We display the total 
    $\chi^2$ for each data set along with the number of observables $n$ and
    the data set's $p-$value.  The set named `quarks' contains $BR(B_s
    \rightarrow \phi \mu^+ \mu^-)$, 
    $BR(B_s\rightarrow \mu^+\mu^-)$,  $\Delta m_s$ and various differential
    distributions in $B\rightarrow K^{(\ast)} \mu^+\mu^-$ decays among others,
    whereas `LFU
    FCNCs' contains $R_{K^{(\ast)}}$ and some $B$ meson decay branching ratios
  into di-taus. {Our sets are identical to those defined by \smelli{}
  and we refer the curious reader to its manual~\cite{Aebischer:2018iyb},
  where the observables are enumerated.}
  We have updated $R_K$ and $BR(B_{s,d}\rightarrow \mu^+ \mu^-)$ with the
    latest LHCb measurements as detailed in \S\ref{sec:fits}.
Two free parameters of the model were
    fitted: $\theta_{23}=\DTFHMtheta{}$ and $g_{X}=\DTFHMgzp{}$.} 
  \end{center}
  \end{table}
The $p-$value is significantly lower than the canonical lower bound of 0.05, indicating a
somewhat poor fit.

\subsection{The $\theta_{23}=0$ limit of our models}
\begin{figure}
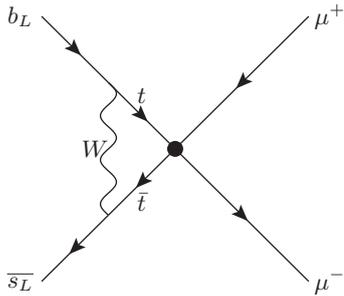

\begin{center}
    \begin{axopicture}(100,100)
      \Line[arrow](100,100)(50,50)
      \Text(108,100)[c]{$\mu^+$}
      \Line[arrow](50,50)(100,0)
      \Text(108,0)[c]{$\mu^-$}
      \Line[arrow](0,100)(25,75)
      \Line[arrow](25,75)(50,50)
      \Line[arrow](50,50)(25,25)
      \Line[arrow](25,25)(0,0)
      \Text(-8,100)[c]{$b_L$}
      \Text(-8,0)[c]{$\overline{s_L}$}
      \Text(37.5,70)[c]{$t$}
      \Text(37.5,30)[c]{$\bar t$}
      \Photon(25,25)(25,75){3}{3}
      \Text(20,50)[c]{$W$}
      \Vertex(50,50){3}
  \end{axopicture}
\end{center}
\caption{\label{fig:loop} Example of the $W-$loop process dominating the SMEFT
  contribution to the
  NCBAs in the $\theta_{23}=0$ case. The filled disc marks the location of the
BSM operator.}
\end{figure}

The NCBAs can receive sizeable contributions even when the tree-level coupling of the $X$ 
boson {to} ${\bar b} s$ {vanishes}. {For example,} non-zero and ${\mathcal
  O}$(1) $\text{TeV}^2$ values for ${C_{lu}}^{2233}$ ({as well as non-zero }
$C_{eu}^{2233}$ {in the case of} the \DTFHM{}  and the \DTFHMp{}) can give a
  reasonable fit to the NCBA data~\cite{Camargo-Molina:2018cwu} via a
  $W$-boson loop as in Fig.~\ref{fig:loop}\footnote{The connection between the EWPOs and minimal flavour-violating one-loop induced
      NCBAs
      has recently been
      addressed in Ref.~\cite{Alasfar:2020mne}, encoding the inputs in the
      SMEFT framework. Third family hypercharge models are in a sense
      orthogonal to this analysis, since they favour
      particular directions in {\em non}-minimal flavour
      violating SMEFT operator space generated already at the {\em tree}-level. The possibility of accommodating the NCBAs through such loop contributions was first pointed out in Ref.~\cite{Becirevic:2017jtw} with a more general study for $b\to s\ell\ell$ transitions presented in Ref.~\cite{Coy:2019rfr}.}
Such a scenario would require that $V_{ts}$ originates
from mixing entirely within the up-quark sector.
{This qualitatively different quark mixing ansatz therefore provides a
motivation to consider the $\theta_{23}=0$ scenario 
    separately.} In the $\theta_{23}=0$ limit, we have that ${\Lambda_{\xi \, 23}^{(d_L)}}$ is
proportional to $s_{13} \ll 1$ meaning that we also predict a negligible
$(C^{(1)}_{lq})^{2223}$ at $M_X$. Meanwhile, $C_{lu}^{2233}$ in the \TFHM{} ({as well as}
$C_{eu}^{2233}$ for the \DTFHMp{} and the \DTFHM{}) remains the same as in the
$\theta_{23}\neq0$ case shown in Tables~\ref{tab:tfhm_wcs}
and~\ref{tab:dtfhmp_wcs}. We note that contributions to the NCBAs arising from 
$W$ loops such as the one in Fig.~\ref{fig:loop} are nevertheless
always included {(through renormalisation group running)} by \smelli{}, even for
$\theta_{23}\neq0$.  

While it is true that {much} of the tension with the NCBAs can be ameliorated by
such $W$ loop contributions, we find {from} our {global} fits that the corresponding values for
$g^2_{X}/M^2_{X}$ are far too large to {simultaneously} give a good fit {to the} EWPOs in {this} $\theta_{23}=0$ {limit}. As our
results with a floating $\theta_{23}$ suggest (see
e.g.~Fig.~\ref{fig:tfhm_global} along $\theta_{23}=0$), {as far as the
EWPOs go}, SM-like scenarios
are strongly preferred, since EWPOs quickly exclude any region that
might resolve the tension with NCBAs. This is even more so than for the
simplified model studied in \cite{Camargo-Molina:2018cwu}, where already a
significant tension with $Z \rightarrow \mu \mu$ was pointed out. In our case,
besides several stringent bounds from other observables measured at the $Z^0$-pole 
for $g_{X} \approx 1$ and $M_{X} \approx O(3) \mbox{~TeV}$ (which predicts
$C_{lq}^{2223}$ just large enough to
give a slightly better fit to $R_K$ and $R_{K^*}$ than the SM) the predicted
$W-$boson mass is more than 5$\sigma$ away from its measured value. This stands
in stark contrast with the overall better-than-SM fit we find for $M_W$ when
$\theta_{23}\neq 0$. 

\section{Discussion \label{sec:disc}}

Previous explorations of {the} parameter spaces of third family hypercharge
models~\cite{Allanach:2018lvl,Allanach:2019iiy} 
{capable of explaining} the neutral current $B-$anomalies
showed {the} 95$\%$ confidence level exclusion regions from various {important} constraints, but {these analyses} did not
include electroweak precision observables. 
Collectively, the electroweak precision observables were potentially a {model-killing constraint} because, through the $Z^0-Z^\prime$
mixing predicted in the models, the prediction of $M_W$ in terms of $M_Z$ is {significantly}
altered from the SM prediction. This was noticed in
Ref.~\cite{Davighi:2020nhv}, where rough estimates of the absolute sizes of
deviations were made. However, {the severity of this constraint on the original Third Family Hypercharge ($Y_3$) Model parameter space was 
found to depend greatly} upon which estimate\footnote{{In more detail, the strength of the constraint was strongly dependent on  how many of the oblique parameters $S$, $T$, and $U$ were allowed to simultaneously float.}} {of the constraint} was used. 
In the present paper, we use the \smelli{}~\cite{Aebischer:2018iyb} computer
program to robustly and accurately predict the electroweak precision
observables and provide a comparison with empirical measurements.
We have {thence carried out} {\em global fits}\/ of third family hypercharge
models to data pertinent to the neutral current
$B-$anomalies as well as the electroweak precision observables. This is more sophisticated than the previous efforts because it
allows tensions between 217 different observables to be traded off against one
another in a statistically sound way. In fact, at the best-fit points of the
third family hypercharge 
models, $M_W$ fits {\em better}\/ than in the SM, whose prediction is some 2$\sigma$
too low.

One ingredient of our fits was the assumption of a fermion mixing
ansatz. The precise details of fermion mixing are expected to be fixed in
third family hypercharge models by a more complete ultra-violet model.
This could lead
to suppressed non-renormalisable operators in the third family hypercharge
model effective field theory, for example which, when the flavon acquires its
VEV, lead to small mixing effects. Such detailed model building seems
premature {in the absence of} additional information coming from the direct observation of
a flavour-violating $Z^\prime$, or indeed independent precise confirmation of
NCBAs from the Belle II~\cite{Kou:2018nap} experiment.
Reining in any urge to delve into the underlying model building, we prefer simply to assume
 a non-trivial structure in the fermion mixing matrix which
changes the observables we consider most: {those involving} the left-handed down
quarks. Since the neutral current $B-$anomalies are most sensitive to the
mixing angle between left-handed bottom and strange quarks, we have allowed {this angle}
to float. But the other mixing angles and complex phase in the
matrix have been set to some
(roughly mandated but ultimately arbitrary) values equal to those in the CKM
matrix. We have checked that changing these arbitrary values somewhat
({{\em e.g.}}\ setting them to zero) does not change the fit qualitatively: a change
in $\chi^2$ of up to 2 units was observed. It is clear that a more thorough
investigation of such variations
may become interesting in the future, particularly if the NCBAs strengthen 
because of new measurements.

\begin{table}
  \begin{center}
    \begin{tabular}{|c|c|c|} \hline
      model & $\chi^2$ & $p-$value \\ \hline
      SM & \SMchi2{} & \SMpval{} \\
      $Y_3$ & \TFHMchi2{} & \TFHMpval{}\\
      $DY_3$ & \DTFHMchi2{} & \DTFHMpval{} \\
      $DY^\prime_3$ & \DTFHMpchi2{} & \DTFHMppval{}\\
      \hline
      \end{tabular}
    \end{center}
    \caption{\label{tab:summary} Comparison of $p-$values {resulting from our} global fits of the SM
      and various third family hypercharge models (with $M_{Z^\prime}=3$ TeV)
      to a combination of $219$ neutral current $B-$anomaly and electroweak data.}
\end{table}
We summarise the punch line of the global fits in Table~\ref{tab:summary}. We
see that, while the SM suffers from a poor fit to the combined data set, the
various third family hypercharge models fare considerably better. The model
with the best fit is the original Third Family Hypercharge Model ($Y_3$).
We have presented the constraints upon the parameter spaces of the \TFHM{} and
the \DTFHMp{} {in detail} in \S\ref{sec:fits}. The qualitative behaviour of the \TFHM{}
and the \DTFHMp{} in the
global fit is similar, although the regions of preferred parameters
are different. 

It is well known that $\Delta m_s$ provides a strong constraint on
  models which fit the NCBAs and ours are no exception: in fact, we see in
  Figs~\ref{fig:tfhm_int},\ref{fig:dtfhmp_int} that this variable has a pull of
  $2.7\sigma$ ($2.1\sigma$) in the 
  \TFHM{} (\DTFHMp{}), whereas the SM pull is only $1.1\sigma$, according to
  the \smelli{} calculation.
  The dominant beyond the SM contribution to $\Delta m_s$ from our models is
  proportional to the $Z^\prime$ coupling to $\bar s b$ quarks squared, i.e.\
  $[g_X (\Lambda_\xi^{(d_L)})_{23} / 6]^2$. The coupling $(\Lambda_\xi^{(d_L)})_{23}$ is adjustable
  because $\theta_{23}$ is allowed to vary over the fit, and $\Delta m_s$
  provides an {\em upper bound}\/ upon $|g_X (\Lambda_\xi^{(d_L)})_{23}|$.
  On the other hand, in
  order to produce a large enough effect in the lepton flavour universality
  violating observables to fit data, the {\em product}\/ of the $Z^\prime$ couplings
  to $\bar s b$ quarks and to $\mu^+ \mu^-$ must be at least a certain
  size.   Thus, models where the $Z^\prime$ couples more strongly to muons because
  their $U(1)_X$ charges are larger fare better when fitting the combination
  of the LFU FCNCs and $\Delta m_s$.
  The $Z^\prime$ coupling to muons is 1/2
  for the \TFHM{} and 2/3 for the \DTFHMp{}, favouring the \DTFHMp{} in
  this regard.
  
 Despite the somewhat worse fit to $\Delta m_s$ for the
  \TFHM{} as compared to the \DTFHMp{}, Table~\ref{tab:summary} shows that,
  overall, the \TFHM{} is 
  a better fit. Looking at the flavour observables in detail, it is hard to
  divine a single cause for this: it appears to be the accumulated effect of
  many flavour observables in tandem. The difference in $\chi^2$
  between the \TFHM{} and the \DTFHMp{} of 4.4 is not large and
  might merely be the result of statistical fluctuations in the 219 data;
  indeed 1.2 of this comes from the difference of quality of fit to the EWPOs.

All of the usual caveats levelled at interpreting $p-$values
apply. In particular, $p-$values change depending upon exactly which
observables are included or excluded. We have stuck to pre-defined sets of observables in
\smelli{} in an attempt to reduce bias. However, we note that there are other
data that are in tension with SM predictions which we have not included, namely
the anomalous magnetic moment of the muon $(g-2)_\mu$ and charged current
$B-$anomalies. If we were to include these observables, the $p-$values of
all models in Table~\ref{tab:summary} would lower. Since third family hypercharge models
give the {\em same}\/ prediction for these observables as the SM, each
model would receive the same $\chi^2$ increase as well as the increase 
in the number of fitted data. However, since our models have essentially nothing to add to these observables compared with the SM, we feel justified in leaving them out from the beginning. We could have excluded some of the observables that \smelli{} includes in our data sets (obvious choices include those that do not involve bottom quarks,
e.g.\ $\epsilon_K$) further changing our calculation of the $p-$values of the
various models. 

{As noted above, as far as the third
  family hypercharge models currently stand,
  the $Z^\prime$ contribution to $(g-2)_\mu$ is small~\cite{Allanach:2015gkd}.
  In order to explain an inferred beyond the SM contribution to $(g-2)_\mu$
  compatible with current measurements $\Delta (g-2)_\mu/2 \approx 28\pm 8 \times 10^{-10}$, one 
  may simply add a heavy (TeV-scale) vector-like lepton representation 
  that couples to the muon 
  and the $Z^\prime$ at one vertex. In that case, a one-loop diagram with the heavy leptons
  and $Z^\prime$ running in the loop is sufficient and is simultaneously compatible with
  the neutral current $B-$anomalies and measurements of $(g-2)_\mu$~\cite{Allanach:2015gkd}.}

Independent corroboration from other experiments
and future $B-$anomaly measurements
are eagerly awaited and, depending upon them, a re-visiting of global fits
to flavour and electroweak data
may
well become desirable. We also note that, since electroweak precision
observables play a key r\^{o}le in our fits, an increase in precision upon them
resulting from LHC or future $e^+e^-$ collider data, could also prove to be of
great utility in testing third family hypercharge models indirectly. Direct
production of the predicted $Z^\prime$~\cite{Allanach:2019iiy,Allanach:2019mfl}
(and a measurement of its couplings) would, along with an observation of {\em
  flavonstrahlung}~\cite{Allanach:2020kss},
ultimately provide a `smoking gun' signal. 

\section*{Acknowledgements}
This work has been partially supported by STFC Consolidated HEP grants
ST/P000681/1 and ST/T000694/1. JECM is supported by the Carl Trygger foundation (grant no. CTS
17:139).
We are indebted to M McCullough
for raising the question of the electroweak precision observables in the
\TFHM{}. We thank other members of the
Cambridge Pheno Working Group for discussions and D Straub and P Stangl for
helpful communications about the nuisance parameters in \smelli{}. 
\bibliographystyle{spphys}
\bibliography{tfhms_global}
\end{document}

%% file: anc/sm_fit.tex
quarks & 221.2 & 167 & .0032  \\
LFU FCNCs & 35.3 & 21 & .026  \\
EWPOs & 35.7 & 31 & .26  \\ \hline
global & 292.2 & 219 & .00068  \\

%% file: anc/tfhm_fit.tex
quarks & 192.3 & 167 & .072  \\
LFU FCNCs & 21.0 & 21 & .34  \\
EWPOs & 36.0 & 31 & .17  \\ \hline
global & 249.3 & 219 & .065  \\

%% file: anc/dtfhmp_fit.tex
quarks & 196.6 & 167 & .047  \\
LFU FCNCs & 20.2 & 21 & .39  \\
EWPOs & 37.2 & 31 & .14  \\ \hline
global & 253.9 & 219 & .044  \\

%% file: anc/dtfhm_fit.tex
quarks & 204.7 & 167 & .019  \\
LFU FCNCs & 20.7 & 21 & .35  \\
EWPOs & 35.0 & 31 & .2  \\ \hline
global & 260.5 & 219 & .023  \\